\documentclass[manuscript,screen,nonacm]{acmart}
\AtBeginDocument{%
  }

\setcopyright{acmlicensed}
\copyrightyear{2026}
\acmYear{2026}

\usepackage{enumitem}
\usepackage{array} 
\usepackage{multirow} 
\usepackage[utf8]{inputenc}
\usepackage[T1]{fontenc}    
\usepackage{hyperref}       
\usepackage{url}            
\usepackage{booktabs}       
\usepackage{amsfonts}       
\usepackage{nicefrac}       
\usepackage{microtype}
\usepackage{comment}
\usepackage{caption}
\usepackage{natbib}
\usepackage{doi}

\usepackage{listings}
\usepackage{microtype}
\usepackage{tikz}
\usetikzlibrary{shapes.geometric, arrows.meta, positioning, backgrounds, calc, shadows.blur, fit}

\definecolor{boxGray}{RGB}{150, 150, 150}
\definecolor{boxWhite}{RGB}{255, 255, 255}

\definecolor{colorFP64}{RGB}{189, 215, 238}
\definecolor{colorFP32}{RGB}{248, 203, 173}
\definecolor{colorFP16}{RGB}{198, 194, 229}
\definecolor{colorE5M2}{RGB}{255, 242, 204}
\definecolor{colorBg}{RGB}{248, 250, 245}
\definecolor{lineGray}{RGB}{80, 80, 80}

\usetikzlibrary{positioning, fit, backgrounds, shadows, backgrounds}
\usetikzlibrary{decorations.pathreplacing}

\definecolor{primary}{HTML}{2C3E50}   
\definecolor{accent}{HTML}{2980B9} 
\definecolor{backendCol}{HTML}{ECF0F1}
\definecolor{workerCol}{HTML}{E8F6F3}  
\definecolor{workerLine}{HTML}{1ABC9C}

\definecolor{procBlue}{RGB}{250, 240, 255}
\definecolor{procStroke}{RGB}{0, 102, 204}
\definecolor{dataGreen}{RGB}{235, 250, 235}
\definecolor{dataStroke}{RGB}{34, 139, 34}
\definecolor{algoRed}{RGB}{255, 235, 235}
\definecolor{algoStroke}{RGB}{204, 0, 0}
\definecolor{colSign}{HTML}{B24A3A} 
\definecolor{colExp}{HTML}{D9B2AC} 
\definecolor{colMant}{HTML}{C9CCD6}
\definecolor{colText}{HTML}{1F2933} 
\definecolor{colBorder}{HTML}{6B7280}

\definecolor{procBlue2}{RGB}{235, 245, 251} 
\definecolor{procLine}{RGB}{52, 152, 219} 
\definecolor{toolFill}{RGB}{230, 240, 230} 
\definecolor{toolLine}{RGB}{46, 204, 113}  
\definecolor{coreFill}{RGB}{253, 235, 208} 
\definecolor{coreLine}{RGB}{230, 126, 34}  
\definecolor{darkText}{RGB}{50, 50, 50}

\definecolor{mygreen}{RGB}{28,172,0} 
\definecolor{mylilas}{RGB}{170,55,241}
\definecolor{codegreen}{rgb}{0,0.6,0}
\definecolor{codemilk}{rgb}{0.5,0.5,0.5}
\definecolor{codegray2}{rgb}{0.9,0.9,0.9}
\definecolor{codepurple}{rgb}{0.58,0,0.82}
\definecolor{backcolour}{rgb}{0.95,0.95,0.92}

\usepackage{color} 
\usepackage{amsmath,amsthm}
\usepackage{algorithm}
\usepackage{algpseudocode}
\usepackage{graphicx} 

\usepackage[caption=false]{subfig}

\definecolor{codemilk}{RGB}{254, 246, 228} 

\lstdefinestyle{bashstyle}{
backgroundcolor=\color{codegray2},   
commentstyle=\color{codegreen},
keywordstyle=\color{magenta},
numberstyle=\tiny\color{codemilk},
stringstyle=\color{codepurple},
basicstyle=\ttfamily\scriptsize,
breakatwhitespace=false,         
breaklines=true,                 
captionpos=b,                    
keepspaces=true,                 
numbers=left,                    
numbersep=5pt,                  
showspaces=false,                
showstringspaces=false,
showtabs=false,                  
tabsize=10
}

\lstdefinestyle{cppstyle}{
    language         = C++,
    backgroundcolor  = \color{gray!22},
    commentstyle     = \color{codegreen},
    keywordstyle     = \color{magenta},
    numberstyle      = \tiny\color{codemilk},
    stringstyle      = \color{codepurple},
    basicstyle       = \ttfamily\scriptsize,
    breakatwhitespace = false,
    breaklines       = true,
    captionpos       = b,
    keepspaces       = true,
    numbers          = left,
    numbersep        = 5pt,
    showspaces       = false,
    showstringspaces = false,
    showtabs         = false,
    tabsize          = 2,
    morekeywords     = {__PROMISE__, PROMISE_CHECK_VAR, PROMISE_CHECK_ARRAY, __PR_1__, __PR_2__, __PR_3__},
    emph             = {half_float::half, flx::floatx},
    emphstyle        = \color{orange}\bfseries
}

\definecolor{milkpink}{HTML}{FDF8F6}
\definecolor{codemilk}{HTML}{A67C52} 
\definecolor{codegreen}{HTML}{228B22}
\definecolor{codepurple}{HTML}{800080}
\lstdefinestyle{jsonstyle}{
    backgroundcolor=\color{milkpink},   
    commentstyle=\color{codegreen},
    keywordstyle=\color{magenta},
    numberstyle=\tiny\color{codemilk},
    stringstyle=\color{codepurple},
    basicstyle=\ttfamily\scriptsize,
    breakatwhitespace=false,         
    breaklines=true,                 
    captionpos=b,                    
    keepspaces=true,                 
    numbers=left,                    
    numbersep=5pt,                  
    showspaces=false,                
    showstringspaces=false,
    showtabs=false,                  
    tabsize=10
}

\lstdefinestyle{txtstyle}{
    language={},                
    morekeywords={},             
    keywordstyle=,             
    basicstyle=\ttfamily\scriptsize,
    backgroundcolor=\color{milkpink},
    numbers=left,          
    numberstyle=\tiny\color{gray},
    breaklines=true,
    breakatwhitespace=false,
    showstringspaces=false,
    keepspaces=true,
    columns=fullflexible,
    emph={for},
    emphstyle=\color{black}\bfseries
}

\begin{document}
\title{Floating-point autotuning with customized precisions}

\author{Xinye Chen}
\orcid{0000-0003-1778-393X}
\email{xinye.chen@lip6.fr}
\affiliation{%
  \institution{Sorbonne Université, CNRS, LIP6}
  \city{Paris}
  \country{France}
}
\affiliation{%
  \institution{China Telecom Cloud Technology Co., Ltd.}
  \city{Beijing}
  \country{China}
}

\author{Fabienne Jézéquel}
\orcid{0000-0002-8782-7566}
\email{fabienne.jezequel@lip6.fr}
\affiliation{
  \institution{Sorbonne Université, CNRS, LIP6}
  \city{Paris}
  \country{France}
}
\affiliation{
  \institution{Université Paris-Panthéon-Assas}
  \city{Paris}
  \country{France}
}

\author{Thibault Hilaire}
\orcid{0000-0001-6365-9695}
\email{thibault.hilaire@lip6.fr}
\affiliation{%
  \institution{Sorbonne Université, CNRS, LIP6}
  \city{Paris}
  \country{France}
 } 
\thanks{This work was supported
 by the NumPEx Exa-MA (ANR-22-EXNU-0002) and the FPT-4 (ANR-24-CE46-7572)
projects of the French National Agency for Research (ANR)  
and performed using HPC resources from GENCI-IDRIS (Grant AD010614986R2). 
}

\renewcommand{\shortauthors}{Chen et al.}

\begin{abstract}
Reduced-precision arithmetic offers significant opportunities to improve performance, memory usage, and energy efficiency in numerical applications, provided that numerical accuracy is preserved. This work investigates automated precision tuning through customized floating-point formats with user-defined exponent and significand sizes, enabling the emulation of emerging low-precision formats and the exploration of non-standard precision configurations within a unified mixed-precision framework. The proposed methodology, implemented in the PROMISE precision autotuning tool, combines numerical validation with a systematic search to generate program variants that satisfy user-defined accuracy requirements. To address the computational cost of this exploration, a containerized benchmarking framework supports parallel execution across multiple algorithms and parameter configurations. The approach is evaluated on a suite of numerical programs, including linear solvers and applications from the Rodinia benchmark. Results show that a substantial proportion of variables can be safely reduced to lower precision while preserving accuracy, indicating that standard double precision is often over-provisioned. These findings highlight the potential of automated precision tuning to derive efficient mixed-precision configurations tailored to application-specific accuracy requirements.
\end{abstract}



\maketitle

\section{Introduction}
\subsection{Precision tuning}
Performing arithmetic with lower precision usually {reduces} computational cost, energy consumption, and communication cost.  With the prevalence of hardware-supported reduced-precision arithmetic, many algorithms can be sped up by hardware accelerators using low-precision arithmetic, as high precision is often unnecessary for every step of their computation. Modern  Graphics Processing Units (GPUs) provide extensive support for low-precision numerical formats. 
Attaining optimal performance in numerical algorithms can often be achieved by quantizing floating-point numbers to low-precision formats while conducting meticulous rounding error analysis to maintain numerical stability and accuracy; see the surveys \citep{Higham_Mary_2022, kashi2025} and references therein.   This solution is critical in scientific computing and high-performance computing (HPC), where resource constraints and numerical stability are important concerns. 

Precision tuning aims to find optimal precision configurations that provide the opportunity of balancing the trade-off between accuracy and speedup. Automated precision tuning typically involves analyzing floating-point operations in a program to assign the lowest precision types that satisfy a numerical accuracy constraint.  A straightforward approach to exploit the advantages of low-precision is to perform a brute-force search over all possible precision configurations. However, this method is not practical for large codebases.  To address this limitation, existing work on automated precision tuning can be broadly categorized into dynamic and static approaches (see \cite{10.1145/3381039} and references therein).   
Static analysis 
 infers precision needs from source code or program structure without execution, mainly using mathematical models or abstract interpretation, 
while dynamic analysis measures numerical behavior during execution to assess and optimize precision usage.  
On top of these, several approaches, such as ADAPT~\citep{Menon2018ADAPT} and CHEF-FP~\citep{Singh2023CHEFFP}, 
leverage automatic differentiation (AD) to estimate program output sensitivity to floating-point precision.

{Here we focus on dynamic analysis tools, which typically include the search-based tools Precimonious~\citep{10.1145/2503210.2503296} and PROMISE~\citep{GRAILLAT2019101017} both
based on the delta debugging algorithm \citep{988498}.}
CRAFT \citep{10.1177/1094342016652462} uses binary instrumentation and
modification to do fine-grained precision analysis.  To improve
precision tuning performance, Blame Analysis~\citep{Rubio2016} uses shadow execution to identify numerically insensitive variables that can be excluded from the search space. Blame Analysis
has successfully been used to decrease Precimonious runtime. 
HiFPTuner \citep{10.1145/3213846.3213862} extends Precimonious by organizing variables into a hierarchical structure
according to their impact on numerical accuracy, enabling a progressive exploration of the
search space that substantially reduces the number of configurations to evaluate. 
{More recently, FPLearner \citep{10548640}, a deep learning-based framework, has been proposed to leverage graph neural networks for performance and accuracy prediction of mixed-precision programs, thereby improving the efficiency of precision tuners such as Precimonious and HiFPTuner.}

{
Our application of interest here focuses on an automated search-based
tool based on delta debugging. 
In \cite{9251260} six search algorithms are compared for precision tuning in a representative set of HPC benchmarks. Only delta debugging (DD) and Genetic Search Algorithm (GA) can identify a valid mixed-precision configuration for all the benchmarks considered, DD producing slightly more performant codes than GA. More recently, through a study of weather and climate models, the effectiveness of automated precision-tuning tools based on delta debugging was validated \citep{10820739}.} 

{As previously mentioned, the precision autotuning tool PROMISE (PRecision OptiMISE)\footnote{https://promise.lip6.fr} \citep{GRAILLAT2019101017,10.1007/978-3-030-72654-6_29} 
leverages the {CADNA library \citep{Jez2008,Eberhart2015HighPerformance}} to
assess numerical accuracy and employs the delta debugging algorithm to find suitable type configurations, currently supporting combinations of FP64, FP32, and FP16 formats.} Unlike conventional precision-tuning tools, PROMISE relies on CADNA instead of a uniform FP64 execution to provide numerically validated reference results, overcoming the limitation that FP64 reference computations may themselves be affected by rounding errors.

\subsection{Contributions}
Because of the emergence of various new low-precision formats and the development of configurable accelerators such as FPGAs, bit-width tuning is crucial to minimize resource usage while preserving accuracy.  In this paper we introduce a new version of PROMISE by extending it to customized low-precision formats, aligning with the current trend toward hardware-efficient numerical computing.  {Compared with existing precision tuners that typically operate over native or predefined precision sets, the new PROMISE version exposes the exponent bitwidth and trailing-significand bitwidth as user-level search parameters.} With this new PROMISE version, we enable systematic exploration of non-standard significand and exponent configurations within a unified mixed-precision framework.

Determining a valid mixed-precision configuration for a given numerical application typically involves running PROMISE across multiple hyper-parameter settings to capture precision variability. A containerized benchmarking tool addresses this requirement by enabling the parallel execution of PROMISE. {Together,} we present our benchmark study for precision tuning in programs addressing scientific computing problems. The main contributions in this article are the following:
\begin{enumerate}
\item  

We propose a new version of PROMISE that extends precision tuning to user-defined floating-point formats, specified by their exponent and trailing-significand bitwidths. These customized formats broaden the search space and enable finer-grained exploitation of reduced-precision arithmetic. To the best of our knowledge, this is the first automated source-to-source precision-tuning framework that can explore user-defined floating-point formats while consistently supporting elementary mathematical functions.

\item We have developed a mixed-precision benchmarking tool for automated precision tuning targeting large-scale deployment.
We also propose a suite of simulated benchmarks for numerical algorithms, providing insights into low-precision configurations without prior knowledge of mixed-precision algorithm design.
\end{enumerate}

We validate our approach  
with extensive simulations on algorithms for numerical approximation and linear system solving. We believe that this study can provide guidance on precision tuning for the scientific community. Our contribution to the PROMISE software is integrated into \url{https://gitlab.lip6.fr/pequan/promise}. Moreover, the public artifact repository at \url{https://github.com/PEQUAN/hpc-mix-bench} provides the original source code alongside the PROMISE-transformed mixed-precision versions corresponding to each reported accuracy target. The rest of the paper is organized as follows: Section~\ref{sec:recap} discusses the foundation and the new features of PROMISE. Section~\ref{sec:bench} details the simulation methods and the benchmark setup, analyzing the 
mixed-precision configurations produced for varying requested accuracies. Finally, Section~\ref{sec:conclude} concludes the paper.

\section{PROMISE: dynamic precision analysis with numerical validation}\label{sec:recap}

\subsection{Discrete Stochastic Arithmetic}
Using a probabilistic approach, the CESTAC method~\citep{VIGNES1993233} 
estimates the propagation of rounding errors in floating-point computations by: (1) executing a given computational code $N$ times, each with a distinct rounding error propagation, and (2) deriving an estimate of the common component across these executions, which serves as a reliable approximation of the exact result. 

To simulate varying rounding error propagations, CESTAC employs a \textit{random rounding mode}: each floating-point result is rounded up or down with the probability 0.5. 
Using the CESTAC method, a set of $N$ computed results $R_i$, $i = 1, \dots, N$, is generated to calculate the mean $\overline{R} = \frac{1}{N} \sum_{i=1}^N R_i$, which is taken as the representative.  
%
Under certain assumptions \citep{VIGNES1993233}, 
the number of exact significant digits in $\overline{R}$, denoted $C_{\overline{R}}$, can be estimated as
$C_{\overline{R}} = \log_{10} \left( \frac{\sqrt{N} |\overline{R}|}{\sigma \tau_{\beta}} \right)$,
where $\sigma = \sqrt{\frac{1}{N-1} \sum_{i=1}^N (R_i - \overline{R})^2}$ is the sample standard deviation, and $\tau_{\beta}$ is the critical value of the Student’s $t$-distribution with $N-1$ degrees of freedom at a confidence level of $1 - \beta$. Typically, $N = 3$, $\beta = 0.05$, and $\tau_{\beta} = 4.43$. Indeed, it has been shown \citep{VIGNES1993233} 
that $N = 3$ balances reliability and computational efficiency. 
However, 
{ 
$C_{\overline{R}}$ may be invalid when multiplication operands or divisors lack significance, requiring dynamic monitoring, termed \textit{self-validation}.
This motivates synchronous computation of the $N$ results $R_i$ and the notion of \textit{computational zero}, representing either true zero or numerical noise. In the latter case,  $C_{\overline{R}} \leq 0$, meaning that $\overline{R}$  has no correct significant digits.
}

Discrete Stochastic Arithmetic (DSA) \citep{Vignes2004} combines the CESTAC method, the notion of computational zero, and discrete stochastic relations. This framework enables the monitoring of scientific computations, the detection of numerical instabilities, and the verification of the underlying assumptions, thereby enhancing the robustness and reliability of numerical results. 
{
The CADNA\footnote{\url{https://cadna.lip6.fr}} library \citep{Jez2008} implements DSA using stochastic arithmetic types and functions. Users replace ordinary floating-point variables with CADNA types, requiring only minor changes to C/C++ code. Alternatively, the CADNAIZER tool automatically transforms existing C/C++ code into CADNA-instrumented code. CADNA then estimates the number of correct significant digits, detects severe numerical instabilities and numerical noise, and can account for input uncertainties.
}

If the numerical accuracy of specific parts of the code proves insufficient, the user can improve them by adopting more numerically robust algorithms or by increasing the arithmetic precision. Conversely, when the required level of accuracy is achieved, the code becomes a suitable candidate for mixed-precision optimization using precision autotuning tools, with the goal of improving computational performance.

\subsection{Automated precision tuning with Discrete Stochastic Arithmetic}

PROMISE is a software framework designed to identify locally optimal mixed-precision configurations for numerical software by integrating DSA, code instrumentation, and automated source-to-source transformations. It operates on C and C++ programs, requiring the user to provide the code under investigation together with annotations specifying precision-tuning candidates and the desired accuracy of the results, typically expressed as a target number of correct significant digits. PROMISE first transforms the input program into an instrumented version using the CADNAIZER tool. This instrumented code is then executed with the CADNA library, which employs DSA to estimate the number of exact significant digits and produce numerically validated reference results. Based on this information, PROMISE explores the space of mixed-precision configurations using a delta debugging algorithm. As illustrated in \figurename~\ref{fig:deltadebugging} (for the case of two precision types), the search starts from a high-precision baseline and progressively lowers precision while discarding configurations that fail to meet the required numerical accuracy. When a configuration satisfies the accuracy criterion, the search continues in the same direction without exploring alternative configurations at the same level, thereby avoiding an exhaustive exploration of the search space. This process identifies a locally minimal set of program locations that must retain higher precision while assigning the lowest possible precision to the remaining operations. Thus, PROMISE derives a locally minimal mixed-precision mapping that maximizes low-precision use while satisfying the requested accuracy. 

\begin{figure}[!t] 
\centering
\begin{minipage}{0.45\linewidth}
    \centering
    \resizebox{\linewidth}{!}{
\begin{tikzpicture}[
    >=Stealth,
    font=\large\sffamily,
    thick,
    text=black
]

\newcommand{\blockFullWhite}[3]{
    \begin{scope}[shift={(#2,#3)}]
        \fill[boxWhite] (-1.2, -0.25) rectangle (1.2, 0.25);
        \draw[thick] (-1.2, -0.25) rectangle (1.2, 0.25);
        \coordinate (#1_bot) at (0, -0.25); \coordinate (#1_top) at (0, 0.25); \coordinate (#1_left) at (-1.2, 0);
    \end{scope}
}

\newcommand{\blockFullGray}[3]{
    \begin{scope}[shift={(#2,#3)}]
        \fill[boxGray] (-1.2, -0.25) rectangle (1.2, 0.25);
        \draw[thick] (-1.2, -0.25) rectangle (1.2, 0.25);
        \coordinate (#1_bot) at (0, -0.25); \coordinate (#1_top) at (0, 0.25); \coordinate (#1_left) at (-1.2, 0);
    \end{scope}
}

\newcommand{\blockHalfLeft}[3]{
    \begin{scope}[shift={(#2,#3)}]
        \fill[boxGray] (-1.2, -0.25) rectangle (0, 0.25);
        \fill[boxWhite] (0, -0.25) rectangle (1.2, 0.25);
        \draw[thick] (-1.2, -0.25) rectangle (1.2, 0.25);
        \draw[thin] (0, -0.25) -- (0, 0.25); 
        \coordinate (#1_bot) at (0, -0.25); \coordinate (#1_top) at (0, 0.25); \coordinate (#1_left) at (-1.2, 0);
    \end{scope}
}

\newcommand{\blockHalfRight}[3]{
    \begin{scope}[shift={(#2,#3)}]
        \fill[boxWhite] (-1.2, -0.25) rectangle (0, 0.25);
        \fill[boxGray] (0, -0.25) rectangle (1.2, 0.25);
        \draw[thick] (-1.2, -0.25) rectangle (1.2, 0.25);
        \draw[thin] (0, -0.25) -- (0, 0.25);
        \coordinate (#1_bot) at (0, -0.25); \coordinate (#1_top) at (0, 0.25); \coordinate (#1_left) at (-1.2, 0);
    \end{scope}
}

\newcommand{\blockQOne}[3]{
    \begin{scope}[shift={(#2,#3)}]
        \fill[boxGray] (-1.2, -0.25) rectangle (-0.6, 0.25);
        \fill[boxWhite] (-0.6, -0.25) rectangle (1.2, 0.25);
        \draw[thick] (-1.2, -0.25) rectangle (1.2, 0.25);
        \draw[thin] (-0.6, -0.25) -- (-0.6, 0.25); 
        \coordinate (#1_bot) at (0, -0.25); \coordinate (#1_top) at (0, 0.25); \coordinate (#1_left) at (-1.2, 0);
    \end{scope}
}

\newcommand{\blockQTwo}[3]{
    \begin{scope}[shift={(#2,#3)}]
        \fill[boxWhite] (-1.2, -0.25) rectangle (-0.6, 0.25);
        \fill[boxGray] (-0.6, -0.25) rectangle (0, 0.25);
        \fill[boxWhite] (0, -0.25) rectangle (1.2, 0.25);
        \draw[thick] (-1.2, -0.25) rectangle (1.2, 0.25);
        \draw[thin] (-0.6, -0.25) -- (-0.6, 0.25); \draw[thin] (0, -0.25) -- (0, 0.25);
        \coordinate (#1_bot) at (0, -0.25); \coordinate (#1_top) at (0, 0.25); \coordinate (#1_left) at (-1.2, 0);
    \end{scope}
}

\newcommand{\blockQThree}[3]{
    \begin{scope}[shift={(#2,#3)}]
        \fill[boxWhite] (-1.2, -0.25) rectangle (0, 0.25);
        \fill[boxGray] (0, -0.25) rectangle (0.6, 0.25);
        \fill[boxWhite] (0.6, -0.25) rectangle (1.2, 0.25);
        \draw[thick] (-1.2, -0.25) rectangle (1.2, 0.25);
        \draw[thin] (0, -0.25) -- (0, 0.25); \draw[thin] (0.6, -0.25) -- (0.6, 0.25);
        \coordinate (#1_bot) at (0, -0.25); \coordinate (#1_top) at (0, 0.25); \coordinate (#1_left) at (-1.2, 0);
    \end{scope}
}

\newcommand{\blockQFour}[3]{
    \begin{scope}[shift={(#2,#3)}]
        \fill[boxWhite] (-1.2, -0.25) rectangle (0.6, 0.25);
        \fill[boxGray] (0.6, -0.25) rectangle (1.2, 0.25);
        \draw[thick] (-1.2, -0.25) rectangle (1.2, 0.25);
        \draw[thin] (0.6, -0.25) -- (0.6, 0.25);
        \coordinate (#1_bot) at (0, -0.25); \coordinate (#1_top) at (0, 0.25); \coordinate (#1_left) at (-1.2, 0);
    \end{scope}
}

\newcommand{\blockQOneThree}[3]{
    \begin{scope}[shift={(#2,#3)}]
        \fill[boxGray] (-1.2, -0.25) rectangle (-0.6, 0.25);
        \fill[boxWhite] (-0.6, -0.25) rectangle (0, 0.25);
        \fill[boxGray] (0, -0.25) rectangle (0.6, 0.25);
        \fill[boxWhite] (0.6, -0.25) rectangle (1.2, 0.25);
        \draw[thick] (-1.2, -0.25) rectangle (1.2, 0.25);
        \draw[thin] (-0.6, -0.25) -- (-0.6, 0.25); \draw[thin] (0, -0.25) -- (0, 0.25); \draw[thin] (0.6, -0.25) -- (0.6, 0.25);
        \coordinate (#1_bot) at (0, -0.25); \coordinate (#1_top) at (0, 0.25); \coordinate (#1_left) at (-1.2, 0);
    \end{scope}
}

\newcommand{\blockQOneFour}[3]{
    \begin{scope}[shift={(#2,#3)}]
        \fill[boxGray] (-1.2, -0.25) rectangle (-0.6, 0.25);
        \fill[boxWhite] (-0.6, -0.25) rectangle (0.6, 0.25);
        \fill[boxGray] (0.6, -0.25) rectangle (1.2, 0.25);
        \draw[thick] (-1.2, -0.25) rectangle (1.2, 0.25);
        \draw[thin] (-0.6, -0.25) -- (-0.6, 0.25); \draw[thin] (0.6, -0.25) -- (0.6, 0.25);
        \coordinate (#1_bot) at (0, -0.25); \coordinate (#1_top) at (0, 0.25); \coordinate (#1_left) at (-1.2, 0);
    \end{scope}
}

\newcommand{\cmark}[2]{ \begin{scope}[shift={(#1,#2)}] \draw[line width=2.5pt] (0,0) -- (0.15,-0.2) -- (0.45,0.35); \end{scope} }
\newcommand{\xmark}[2]{ \begin{scope}[shift={(#1,#2)}] \draw[line width=2.5pt] (-0.2,-0.2) -- (0.2,0.2); \draw[line width=2.5pt] (-0.2,0.2) -- (0.2,-0.2); \end{scope} }

\blockFullWhite{N0}{0}{0}
\blockFullGray{N1}{0}{-1.8}
\blockHalfLeft{N2L}{-2.5}{-3.6}
\blockHalfRight{N2R}{2.5}{-3.6}
\blockQOne{N3A}{-4.5}{-5.4}
\blockQTwo{N3B}{-1.5}{-5.4}
\blockQThree{N3C}{1.5}{-5.4}
\blockQFour{N3D}{4.5}{-5.4}
\blockHalfLeft{N4A}{-7.0}{-7.6}      
\blockQOneThree{N4B}{-4.5}{-7.6}     
\blockQOneFour{N4C}{-1.5}{-7.6}

\draw[->, thick] (N0_bot) -- (N1_top);

\draw[->, thick] (N1_bot) -- (N2L_top);
\draw[->, thick] (N1_bot) -- (N2R_top);

\draw[->, thick, bend right=22] (N1_bot) to ([xshift=-0.3cm]N3A_top);
\draw[->, thick, bend left=22]  (N1_bot) to ([xshift=0.3cm]N3D_top);

\draw[->, thick] (N2L_bot) -- (N3A_top);
\draw[->, thick] (N2L_bot) -- (N3B_top);

\draw[->, thick] (N2R_bot) -- (N3C_top);
\draw[->, thick] (N2R_bot) -- (N3D_top);

\draw[->, thick] (N3A_bot) -- (N4A_top);
\draw[->, thick] (N3A_bot) -- (N4B_top);
\draw[->, thick] (N3A_bot) -- (N4C_top);

\draw[->, thick] (N4B_bot) -- ++(0, -0.8);
\node at ($(N4B_bot) + (0, -1.2)$) {\Huge $\dots$};

\cmark{1.6}{-0.2}     

\xmark{1.6}{-2.0}     
\xmark{-1.0}{-3.8}    
\xmark{1.0}{-3.8}     

\cmark{-6.2}{-5.5}    

\cmark{-4.0}{-8.4}

\node[anchor=east] (L_High) at (-2.0, 0) {Higher precision};
\draw[thick] (L_High.east) -- (N0_left);

\node[anchor=east] (L_Low)  at (-2.0, -1.8) {Lower precision};
\draw[thick] (L_Low.east) -- (N1_left);

\draw[thick, dashed] (-7.0, -7.6) ellipse (1.6cm and 0.7cm);
\node at (-7.0, -8.7) {Already tested};

\draw[thick, dashed] (-3.1, -4.5) rectangle (6.0, -8.2);
\node at (1.5, -7.2) {Not tested};

\end{tikzpicture}
    }
    \caption{Delta debugging for two precisions (Figure from~\cite{GRAILLAT2019101017}).}
    \label{fig:deltadebugging}
\end{minipage}%
\hfill \hfill
\begin{minipage}{0.51\linewidth}
    \centering
    \resizebox{\linewidth}{!}{
        \begin{tikzpicture}[
            >=Stealth, 
            font=\large\sffamily, 
            thick, 
            text=black!90
        ]

            \def\leftX{1.2}  
            \def\busX{3.2}   
            \def\rightX{4.2} 

            \node (init) [align=center] at (\leftX, 0) {Initial code};
            \node (inst) [below=0.8cm of init] {Instrumented code};
            \node (cadna) [draw, rectangle, minimum width=2cm, minimum height=0.8cm, fill=white, below=0.8cm of inst] {CADNA};
            \node (ref) [below=0.8cm of cadna] {Reference};

            \draw[->, lineGray] (init) -- (inst);
            \draw[->, lineGray] (inst) -- (cadna);
            \draw[->, lineGray] (cadna) -- (ref);

            \coordinate (busTop) at (\busX, 0.15);
            \coordinate (busBot) at (\busX, -4.35);
            \draw[lineGray] (busTop) -- (busBot);
            \draw[lineGray] (inst.east) -- (inst.east -| busTop);
            \draw[lineGray] (ref.east) -- (ref.east -| busBot);
            \node[text=red!80!black, font=\footnotesize\itshape, fill=white, inner sep=2pt] at (\busX, -2.2) {Comparison};
            \draw[->, lineGray] (\busX, 0.15) -- (\rightX, 0.15);
            \draw[->, lineGray] (\busX, -1.35) -- (\rightX, -1.35);
            \draw[->, lineGray] (\busX, -2.85) -- (\rightX, -2.85);
            \draw[->, lineGray] (\busX, -4.35) -- (\rightX, -4.35);

            \begin{scope}[shift={(\rightX, 0.5)}] 
                \node[anchor=west, font=\bfseries\large] (title) at (0, 0.8) {Delta Debug (DD) $\rightarrow$ Decreasing precision format}; 
                
                \fill[colorFP64] (0, 0) rectangle (8, -0.7);

                \fill[colorFP64] (0, -1.5) rectangle (4, -2.2);
                \fill[colorFP32] (4, -1.5) rectangle (8, -2.2);
                \draw[->, blue!60!black] (4, -0.7) -- node[right, black, font=\small] {DD} (4, -1.5);

                \fill[colorFP64] (0, -3.0) rectangle (4, -3.7);
                \fill[colorFP32] (4, -3.0) rectangle (6, -3.7);
                \fill[colorFP16] (6, -3.0) rectangle (8, -3.7);
                \draw[->, blue!60!black] (6, -2.2) -- node[right, black, font=\small] {DD} (6, -3.0);

                \fill[colorFP64] (0, -4.5) rectangle (4, -5.2);
                \fill[colorFP32] (4, -4.5) rectangle (6, -5.2);
                \fill[colorFP16] (6, -4.5) rectangle (7, -5.2);
                \fill[colorE5M2] (7, -4.5) rectangle (8, -5.2);
                \draw[densely dotted] (7, -4.5) rectangle (8, -5.2); 
                \draw[->, blue!60!black] (7, -3.7) -- node[right, black, font=\small] {DD} (7, -4.5);

                \node at (4, -5.6) {\dots\dots};

                \begin{scope}[on background layer]
                    \node[fill=colorBg, inner sep=12pt, rounded corners=4pt, fit={(title) (0,-5.8) (8,0)}] (box) {};
                \end{scope}
            \end{scope}

            \draw[->, lineGray] (\rightX + 8.0, -2.2) -- ++(0.8, 0);
            \node[font=\bfseries, align=center] at (\rightX + 9.2, -2.2) {\rotatebox{-90}{Mixed-precision}\\\rotatebox{-90}{code}};

            \begin{scope}[shift={(5.0, -6.5)}] 
                \foreach \c/\t/\x in {colorFP64/FP64/0, colorFP32/FP32/2, colorFP16/FP16/4} {
                    \fill[\c] (\x, 0) rectangle ++(0.6, 0.4);
                    \node[right, font=\small] at (\x+0.6, 0.2) {\t};
                }
                \fill[colorE5M2] (6, 0) rectangle ++(0.6, 0.4);
                \draw[densely dotted] (6, 0) rectangle ++(0.6, 0.4);
                \node[right, font=\small] at (6.6, 0.2) {E5M2};
            \end{scope}

        \end{tikzpicture}%
    }
    \caption{PROMISE workflow for multiple precisions (the top-to-bottom arrow indicates delta debugging execution).}
    \label{fig:workflow-promise}
\end{minipage}
\end{figure}


\subsection{Customized precision with FloatX and its extension to mathematical functions}
Most hardware implementations adhere to the IEEE 754 standard \citep{ieee754}, which has fixed allocations for sign, exponent, and significand bits. However, some hardware and software ecosystems allow for flexible or custom precision configurations, particularly in domains such as machine learning and scientific computing, as well as in embedded systems. A typical example is Field-Programmable Gate Arrays (FPGAs) which support customizable floating-point precisions. Users can define specific bit allocations for the significand and the exponent to optimize for performance, resource utilization, and energy efficiency. 

FloatX (Float eXtended\footnote{https://github.com/oprecomp/FloatX}) \citep{10.1145/3368086} is a C++ header-only library designed to emulate reduced-precision floating-point formats with user-specific bit-lengths for the exponent and the significand, and to facilitate incremental code transformation. 
{\figurename~\ref{fig:customized_format} illustrates representative examples of types defined in  
FloatX by specifying exponent and significand sizes.} 
{In contrast, GNU MPFR~\citep{10.1145/1236463.1236468}
allows users to control the significand precision, whereas the exponent range is governed by MPFR's exponent type and its global exponent bounds rather than by a user-selected exponent bit field comparable to FloatX.} To preserve efficiency, FloatX leverages heavy inlining and relies on hardware-supported native types (FP32 or FP64) as backends. This design typically achieves higher performance than MPFR in low-precision scenarios, while incurring no additional storage overhead~\citep{10.1145/3368086}. 

\begin{center}
\begin{minipage}{1\linewidth} 
\centering
\resizebox{0.81\linewidth}{!}{
\begin{tikzpicture}[
    font=\large\sffamily, 
    node distance=1.2cm, 
    bitbox/.style={
        draw=colBorder,
        line width=0.5pt,
        minimum height=0.65cm,
        minimum width=0.45cm,
        inner sep=0pt,
        anchor=west
    },
    labelnode/.style={
        anchor=east,
        text=colText,
        align=right,
        font=\large\sffamily, 
        text width=9.5cm 
    },
    legendbrace/.style={
        decorate,
        decoration={brace, mirror, amplitude=4pt, raise=4pt},
        draw=colBorder,
        line width=0.8pt
    },
    legendlabel/.style={
        midway,
        below=8pt,
        font=\sffamily\bfseries, 
        text=colText
    },
    databrace/.style={
        decorate,
        decoration={brace, mirror, amplitude=4pt, raise=12pt},
        draw=colBorder,
        line width=0.8pt
    },
    datalabel/.style={
        midway,
        below=16pt,
        font=\sffamily\bfseries, 
        text=colText
    }
]

    \newcommand{\drawRow}[7]{
        \node[labelnode] (#1_label) at (0, 0) [below=#2] {
            $\texttt{flx::floatx}\langle #3 \rangle \equiv \text{#4}$
        };
        
        \coordinate (start_pos) at ($( #1_label.east) + (0.3cm, 0)$);
        
        \node[bitbox, fill=colSign] (s) at (start_pos) {};
        \coordinate (sign_end) at (s.east);
        \draw [databrace] (start_pos) -- (sign_end) node [datalabel] {1}; 
        
        \coordinate (exp_start) at (s.east);
        \ifnum #7=1
            \foreach \i in {1,...,6} { \node[bitbox, fill=colExp, right=0pt of s] (e\i) {}; \coordinate (s) at (e\i.east); }
            \node[bitbox, fill=colExp, right=0pt of s] (e_last) {}; \coordinate (s) at (e_last.east);
        \else
            \foreach \i in {1,...,#5} { \node[bitbox, fill=colExp, right=0pt of s] (e\i) {}; \coordinate (s) at (e\i.east); }
        \fi
        \coordinate (exp_end) at (s.east);
        \draw [databrace] (exp_start) -- (exp_end) node [datalabel] {#5};
        
        \coordinate (mant_start) at (s.east);
        \ifnum #7=1
            \foreach \i in {1,...,6} { \node[bitbox, fill=colMant, right=0pt of s] (m\i) {}; \coordinate (s) at (m\i.east); }
            \node[right=2pt of s, inner sep=1pt] (break) {$\cdots$};
            \coordinate (s) at ($(break.east) + (2pt,0)$);
            \foreach \i in {1,...,3} { \node[bitbox, fill=colMant, right=0pt of s] (m_end\i) {}; \coordinate (s) at (m_end\i.east); }
        \else
            \foreach \i in {1,...,#6} { \node[bitbox, fill=colMant, right=0pt of s] (m\i) {}; \coordinate (s) at (m\i.east); }
        \fi
        \coordinate (mant_end) at (s.east);
        \draw [databrace] (mant_start) -- (mant_end) node [datalabel] {#6};
    }
    
    \node[labelnode] (l0) at (0,0) {$\mathbf{\texttt{flx::floatx\quad}\langle e,t \rangle \equiv 1+e+t \text{ bits}}$};
    \coordinate (start0) at ($(l0.east) + (0.3cm, 0)$);
    
    \node[bitbox, fill=colSign, minimum width=1.5cm] (legS) at (start0) {};
    \node[font=\bfseries, text=colText] at (legS.center) {Sign};
    \draw [legendbrace] (legS.south west) -- (legS.south east) node [legendlabel] {1};
    
    \node[bitbox, fill=colExp, minimum width=3.5cm, right=0pt of legS] (legE) {};
    \node[font=\bfseries, text=colText] at (legE.center) {Biased Exponent};
    \draw [legendbrace] (legE.south west) -- (legE.south east) node [legendlabel] {e};
    
    \node[bitbox, fill=colMant, minimum width=5.0cm, right=0pt of legE] (legM) {};
    \node[font=\bfseries, text=colText] at (legM.center) {Trailing Significand};
    \draw [legendbrace] (legM.south west) -- (legM.south east) node [legendlabel] {t};
    
    \drawRow{row1}{1.2cm of l0.south}{8, 7}{BF16 (16-bit)}{8}{7}{0}
    \drawRow{row2}{1.2cm of row1_label.south}{5, 2}{E5M2 (8-bit)}{5}{2}{0}
    \drawRow{row3}{1.2cm of row2_label.south}{4, 3}{E4M3 (8-bit)}{4}{3}{0}
\end{tikzpicture}}
\captionof{figure}{{Representative examples of emulated and customized floating-point formats in FloatX.}}
\label{fig:customized_format}
\end{minipage}
\end{center}

Thanks to successive  developments~\citep{GRAILLAT2019101017} \cite{10.1007/978-3-030-72654-6_29},
the current version of PROMISE enables floating-point autotuning using
FP64, FP32 and FP16 formats. As a remark, 
the FP16 format is emulated thanks to a C++ header-only library 
(half.hpp\footnote{https://half.sourceforge.net/}) adhering to the IEEE 754 standard. 
However, PROMISE lacks  support for other low-precision formats (e.g., BF16) as well as customized formats.  Due to the appealing simplicity of using {FloatX} and its above-mentioned features, we integrate FloatX with PROMISE to provide customizable low-precision floating-point support. 
Thanks to FloatX, PROMISE enables users to efficiently exploit reduced-precision in their applications without incurring the overhead of extended-precision formats. 
{
The current PROMISE/FloatX integration emulates each selected format as the program-level numerical format of the corresponding variables and arithmetic operations. 
It does not model hardware execution modes in which data are stored in a low-precision format, such as BF16, while arithmetic is accumulated internally in FP32. Such storage-computation separation would require an additional representation in the transformed code and is left for future work.}
In the new PROMISE version, we keep the half precision library 
that provides the FP16 format, since it performs FP16 arithmetic 
faster than FloatX in a large-scale setting. 
{
However, as the \texttt{half.hpp} library rejects some implicit conversions that appear in transformed benchmark codes, it may stop the delta debugging search before a numerical configuration can be evaluated. Therefore, although \texttt{half.hpp} is retained as an efficient FP16 option, the numerical experiments involving FP16 emulation 
in this article use FloatX to ensure a uniform transformed code path.
 }

We develop a multi-level strategy in PROMISE to explore multiple precision types. 
The method first sorts the types by decreasing precision, then applies the delta debugging algorithm iteratively on consecutive pairs, as illustrated in \figurename~\ref{fig:workflow-promise}. 
{Writing a custom format as $p=(e,t)$, with exponent bitwidth $e$ and trailing-significand bitwidth $t$, PROMISE uses the lexicographic order induced by $(-t,-e)$: larger $t$ gives smaller unit roundoff and is considered first, while larger $e$ breaks ties between formats with the same $t$. For an ordered precision chain $\texttt{p}_1 \succ \texttt{p}_2 \succ \texttt{p}_3 \succ \texttt{p}_4$, where $\succ$ denotes higher precision, PROMISE applies delta debugging successively to adjacent pairs $(\texttt{p}_k,\texttt{p}_{k+1})$; variables validated at $\texttt{p}_{k+1}$ become the candidate set for the next reduction step.}

{To enable practical and realistic reduced-precision emulation of mathematical functions, we extend FloatX with custom-precision implementations based on 
C++ function overloading. 
This support ensures that mathematical functions applied to PROMISE-selected variables are evaluated and returned in the selected precision; otherwise, the function would be evaluated and returned in native precision, breaking the intended custom-precision execution path. 

The current implementation covers the elementary functions used by our benchmarks, including trigonometric functions, exponentials, logarithms, square roots, and powers. For each overloaded function, the input arguments are first rounded to the target FloatX format, the function is evaluated by the corresponding native C/C++ mathematical function in the backend type, and the result is rounded back to the target format, as detailed in Algorithm~\ref{alg:lowprecision}.}

{Specifically, this implementation is conceptually related to low-precision emulation libraries such as \texttt{CPFloat}~\citep{Fasi2023CPFloat} in C and \texttt{pychop} \citep{carson2025} in Python, which also simulate mathematical functions by rounding values, including results of elementary functions, to a target format. We use FloatX as the primary emulation backend because PROMISE performs source-to-source transformations on C/C++ programs and benefits from FloatX's C++ templated types and overloaded arithmetic. In contrast, the C interface of CPFloat is based on explicit function calls, which would require more intrusive rewriting of arithmetic expressions and elementary-function calls in transformed programs.  Our contribution is not a new standalone arithmetic simulator; rather, we extend PROMISE's source-to-source autotuning workflow with FloatX-based emulation of user-defined floating-point formats, enabling these formats to be searched as candidate precision types during delta debugging.  The present FloatX-based implementation emulates target formats whose precision and exponent range fit within the selected native backend. Its rounding accuracy is limited to that of the backend evaluation.}

\begin{algorithm}[ht]
\caption{Reduced-Precision Function Evaluation}
\label{alg:lowprecision}
\begin{algorithmic}[1]
\Require $\theta_1, \ldots, \theta_{\zeta}$: input values;
$f$: function $f: \mathbb{R}^{\zeta} \to \mathbb{R}$;
\Statex \hspace{2pt} \hspace{\algorithmicindent}
$(e, t)$: exponent bits $e$ and significand bits $t$
\For{$i = 1$ to $\zeta$} \Comment{Round each input to the nearest}
  \State $\tilde{\theta}_i \gets \text{round}_{\mathrm{RN}}(\theta_i, e, t)$
\EndFor
\State $y \gets \text{round}_{\mathrm{RN}}(f(\tilde{\theta}_1, \ldots, \tilde{\theta}_{\zeta}), e, t)$
\Comment{Evaluate $f$ and round the output to the nearest}
\State \Return $y$
\end{algorithmic}
\end{algorithm}

\subsection{Configuration and deployment}
\begin{figure}[htp]
\begin{center}    \includegraphics[width=0.9\linewidth]{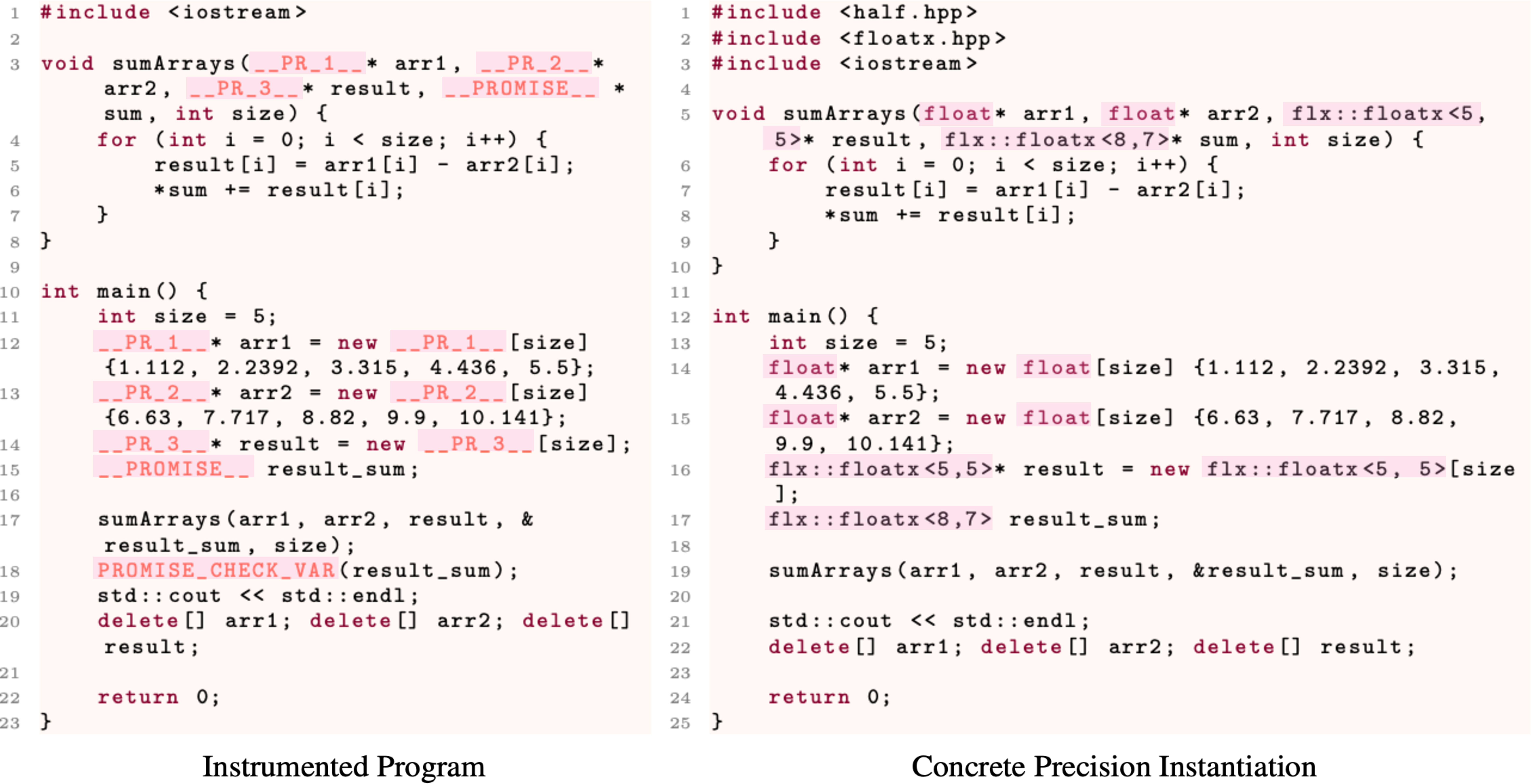}
\end{center}
\caption{Instrumented program and precision instantiation}\label{ftg:inst_ins}
\end{figure}

To enable precision analysis with the PROMISE software, users instrument their code by marking variables for low-precision evaluation using the macros \texttt{\_\_PROMISE\_\_} or \texttt{\_\_PR\_xxx\_\_}. The \texttt{\_\_PROMISE\_\_}  macro  designates a floating-point variable as subject to PROMISE’s precision search. In contrast, \texttt{\_\_PR\_xxx\_\_} macros allow users to group variables so that all variables sharing the same ``\texttt{xxx}'' identifier are assigned the same precision candidate during the mixed-precision exploration: 
each distinct \_\_PR\_xxx\_\_ tag defines its own independent type constraint. These bindings help control precision dependencies and reduce the combinatorial search space. To specify variables or arrays for accuracy validation,  we use \texttt{PROMISE\_CHECK\_VAR} or \texttt{PROMISE\_CHECK\_ARRAY} to check that a variable or all the elements in an array fulfill the accuracy requirement.   
\figurename~\ref{ftg:inst_ins} presents
 an example of instrumented C++ program where  PROMISE\_CHECK\_VAR is used 
for the sum of the element-wise difference between two arrays. 
The tags \_\_PR\_1\_\_, \_\_PR\_2\_\_, and \_\_PR\_3\_\_ 
create separate groups of linked types, e.g., all variables marked \_\_PR\_1\_\_ must share the same type. 
{
The integration of FloatX into PROMISE extends the set of precision formats handled by the source-to-source transformation. During this process, 
\texttt{\_\_PROMISE\_\_} and \texttt{\_\_PR\_x\_\_} annotations are replaced 
by either native types or FloatX 
\texttt{flx::floatx<e,t>} types, and the required FloatX headers are automatically inserted into the generated source code. 
This integration preserves PROMISE’s delta-debugging validation procedure while extending the search space from a fixed set of native formats to user-defined FloatX formats.
}

\begin{figure}[htp]
\captionsetup[lstlisting]{labelformat=empty}
\begin{minipage}{0.25\textwidth}

\begin{lstlisting}[xleftmargin=5.3em, xrightmargin=2.6em,style=jsonstyle]
{"c": [5, 5],
 "b": [8, 7],
 "u": [5, 10]
}
\end{lstlisting}
\captionof{lstlisting}{fp.json} \label{lst:json}
\end{minipage}
\hfill
\begin{minipage}{0.7\textwidth} 
\centering
\begin{lstlisting}[xrightmargin=1.em, style=txtstyle]
compile:
- g++ toy.cpp -O3 -frounding-math -o toy.out -lcadnaC -L$CADNA_PATH/lib -I$CADNA_PATH/include
run: toy.out
files: toy.cpp
log: toy.log
output: debug/
\end{lstlisting}
\captionof{lstlisting}{PROMISE.yml} \label{lst:bash}
\end{minipage}
\caption{Customized-precision and compilation configuration files.}
\label{fig:customized_files}
\end{figure}

The deployment of PROMISE relies on two configuration files, examples
of which are shown in \figurename~\ref{fig:customized_files}. 
In the 
\texttt{promise.yml} file, \texttt{compile} defines the command to compile the source code with CADNA support and optional OpenMP multithreading capabilities, \texttt{run} specifies the executable, \texttt{files} lists source files (default: all \texttt{.cc} files), \texttt{log} designates the log file (optional), and \texttt{output} sets the directory for the transformed code. 
{The \texttt{fp.json} file allows users to define custom floating-point precisions by specifying the exponent and trailing-significand bit-lengths.}
{Once the setup is complete, the requested accuracy is supplied on the command line through \texttt{--nbDigits}; for example, \texttt{promise --nbDigits=3 --precs=cbusd} asks PROMISE to search configurations whose checked outputs share at least three correct 
digits with the CADNA-validated reference.}
The letters \texttt{c}, \texttt{b}, and \texttt{u}
are defined as precision aliases in \texttt{fp.json}.
In this example, \texttt{c} corresponds to a format of 5 exponent bits and 5 significand bits, and \texttt{b} represents BF16, enabling flexible precision tuning for diverse hardware, such as FPGAs. 
The letters \texttt{s} and \texttt{d} denote FP32 and FP64 precision, respectively, and require no further specification. 

\section{{Experimental results}}\label{sec:bench}
In the following, we introduce benchmarking details and describe how to perform PROMISE within them.  All CPU tests were performed on a DELL PowerEdge R750xa server with 2 TB of memory, equipped with two Intel Xeon Gold 6330 processors, featuring 56 cores and 112 threads at 2.00 GHz.  For various programs from the benchmark, we present the resulting precision configurations, taking into account the required accuracy.  

For the CPU results in this section, each PROMISE task is compiled and executed according to its benchmark-specific \texttt{promise.yml} file. The compilation commands use \texttt{g++} version 15.2.0,  the CADNA include/library paths, and the benchmark-specific optimization or OpenMP options when present. The transformed programs remain C/C++ codes, with custom FloatX formats represented as C++ types using operator overloading; therefore, we do not assume that operations on custom formats are vectorized in the same way as native FP32/FP64 SIMD operations. 

\subsection{Benchmark setup}\label{sec:benchmark_setup}

We provide a benchmarking framework that enables the parallel execution of PROMISE across multiple algorithms and parameter configurations. Since each invocation independently explores precision configurations for a given kernel, the workflow is inherently embarrassingly parallel and well-suited for task-level execution. This is particularly important for large benchmark suites, where sequential execution across many algorithms becomes prohibitively time-consuming due to repeated kernel executions during the search and validation phases. To address this, our containerized approach distributes workloads across combinations of algorithms and parameter settings, enabling fine-grained parallelism while avoiding resource contention through controlled job scheduling and ensuring experimental consistency. 
Our approach wraps the PROMISE workflow in a parallelized pipeline that supports two execution modes:
{(i) a Bash-based orchestration using GNU Parallel\footnote{https://www.gnu.org/software/parallel} for lightweight task-level parallelism on a local machine, and (ii) an MPI-based Python scheduler for scalable execution and distributed deployment. The GNU Parallel backend is useful when the benchmark suite is run on one workstation or one shared-memory node because it has minimal setup overhead. The MPI backend is useful when the same independent PROMISE tasks must be distributed across several nodes. Both modes allow the concurrent execution of multiple algorithm kernels.}

The framework scales with available computing resources, permitting efficient utilization of multi-core CPUs and cluster nodes for large benchmark suites. Additionally, the task-level isolation improves robustness, as possible failures in individual PROMISE runs do not affect the execution of other benchmarking tasks. {Such failures may include compilation errors in a transformed source file, nonzero program exits, timeouts, overflow or invalid arithmetic in a candidate custom format,  and configurations that fail the requested digit threshold.} Furthermore,  this workflow is deployed within a preconfigured Docker container to eliminate the overhead of environment setup while preserving consistent performance measures.  The Docker-based deployment ensures reproducibility by encapsulating compiler versions, numerical libraries, and PROMISE dependencies, thereby eliminating variability due to system-level differences.

The schematic of the proposed benchmarking workflow is shown in \figurename~\ref{fig:bench_frame}.  In the reported setup, we launched 6 worker processes to exploit process-level parallelism while avoiding thread oversubscription and memory contention. The value 6 is an experimental scheduling choice for this machine and benchmark suite, not an algorithmic limitation of the framework. Each worker executes one independent PROMISE task; no manual CPU-core pinning is imposed in the reported runs, and the operating system schedules the worker processes on the available cores.  With this setup, the total wall-clock time for the benchmark suite drops from about 13 hours to 3 hours, resulting in a roughly 3.4$\times$ speedup. 
The MPI-based scheduler{—}as an alternative backend{—}remains comparable to GNU Parallel, showing that the scheduling overhead is low. As a result, PROMISE can be used efficiently for large-scale benchmarking, and is particularly valuable for developers seeking mixed-precision variants of their applications across different parameter settings.

\begin{figure}[htp]
\centering
\resizebox{0.62\linewidth}{!}{
\begin{tikzpicture}[
    node distance=1.2cm and 1.2cm,
    font=\sffamily,
    base/.style={
        draw, rounded corners=3pt, thick, align=center,
        inner sep=8pt, drop shadow={opacity=0.1}
    },
    process/.style={
        base, draw=primary, fill=primary!5, text=primary,
        minimum width=3.8cm, minimum height=1.1cm, font=\sffamily\bfseries
    },
    gen/.style={
        base, draw=accent, fill=accent!10, text=accent!80!black,
        minimum width=8cm, minimum height=1.2cm, font=\sffamily\bfseries
    },
    backend/.style={
        base, draw=gray!50, fill=white,
        minimum width=5.5cm, minimum height=2cm
    },
    worker/.style={
        base, draw=workerLine, fill=workerCol,
        minimum width=2.6cm, minimum height=1.6cm, font=\small
    },
    dots/.style={
        minimum width=1cm, text=gray!60, font=\Huge
    },
    arrow/.style={
        -{Stealth[scale=1.2]}, thick, draw=primary!60
    },
    bus/.style={
        draw=primary!30, line width=3pt
    }
]

\node[process] (config) {User Configuration\\ \footnotesize (eg., Target Accuracy - NbDigits)};
\node[process, right=1.2cm of config] (benchmarks) {Benchmark Collection\\ \footnotesize (Benchmark $1 \dots N$)};

\coordinate (topCenter) at ($(config.east)!0.5!(benchmarks.west)$);

\node[gen, below=1.2cm of topCenter] (taskgen) {
    Task Generation \& Decomposition\\
    \footnotesize (Program, Hyperparameter Configuration) Pairs
};

\draw[arrow] (config.south) -- (config.south |- taskgen.north);
\draw[arrow] (benchmarks.south) -- (benchmarks.south |- taskgen.north);

\node[backend, below left=1.5cm and 0.2cm of taskgen.south, anchor=north east] (backendA) {
    \textbf{Execution Backend A}\\
    \tikz\draw[gray!30, thick] (0,0) -- (4.5,0);\\
    \footnotesize Bash + \textsc{GNU Parallel}\\
    \scriptsize \textit{Local Parallelism}
};

\node[backend, below right=1.5cm and 0.2cm of taskgen.south, anchor=north west] (backendB) {
    \textbf{Execution Backend B}\\
    \tikz\draw[gray!30, thick] (0,0) -- (4.5,0);\\
    \footnotesize MPI-based Python Scheduler\\
    \scriptsize \textit{Distributed Execution}
};

\draw[arrow] (taskgen.south) -- ++(0,-0.4) -| (backendA.north);
\draw[arrow] (taskgen.south) -- ++(0,-0.4) -| (backendB.north);

\coordinate (busCenter) at ($(backendA.south)!0.5!(backendB.south) + (0,-1.0)$);

\node[worker] (worker2) at ($(busCenter) + (-1.6cm, -1.2cm)$) {
    \textbf{Task Worker 2}\\
    \rule{1.8cm}{0.2pt}\\
    \scriptsize Task $2 \to$ \textit{PROMISE}
};

\node[worker] (worker1) at ($(worker2) + (-3.2cm, 0)$) {
    \textbf{Task Worker 1}\\
    \rule{1.8cm}{0.2pt}\\
    \scriptsize Task $1 \to$ \textit{PROMISE}
};

\node[worker] (worker3) at ($(busCenter) + (1.6cm, -1.2cm)$) {
    \textbf{Task Worker 3}\\
    \rule{1.8cm}{0.2pt}\\
    \scriptsize Task $3 \to$ \textit{PROMISE}
};

\node[worker] (workerK) at ($(worker3) + (3.6cm, 0)$) {
    \textbf{Task Worker $K$}\\
    \rule{1.8cm}{0.2pt}\\
    \scriptsize Task $K \to$ \textit{PROMISE}
};

\node[dots] at ($(worker3.east)!0.55!(workerK.west)$) {$\dots$};

\draw[bus] (worker1.north |- busCenter) -- (workerK.north |- busCenter);
\draw[thick, primary!30] (backendA.south) -- (backendA.south |- busCenter);
\draw[thick, primary!30] (backendB.south) -- (backendB.south |- busCenter);

\foreach \w in {worker1, worker2, worker3, workerK} {
    \draw[arrow] (\w.north |- busCenter) -- (\w.north);
}

\node[process, below=3.8cm of busCenter, minimum width=12cm, fill=primary!80, text=white] (output) 
    {Aggregated Outputs \& Performance Statistics};

\foreach \w in {worker1, worker2, worker3, workerK} {
    \draw[arrow] (\w.south) -- (\w.south |- output.north);
}

\node[draw=gray!40, dashed, ultra thick, rounded corners=8pt, inner sep=12pt, 
      fit=(backendA)(backendB), 
      label={[anchor=south, font=\bfseries\color{gray!80}]north:Orchestration Layer}] (box1) {};

\node[draw=workerLine!60, dashed, ultra thick, rounded corners=8pt, inner sep=20pt, 
      fit=(worker1)(workerK), 
      label={[anchor=north, font=\bfseries\color{workerLine}]south:Parallel Execution Pool ($K$ Workers)}] (box2) {};
\end{tikzpicture}
}
\caption{{Overview of the benchmarking workflow. User-defined accuracy and precision settings are used to generate fine-grained benchmarking tasks, which are orchestrated and dispatched to a parallel execution pool of $K$ workers. Tasks can be executed either locally using GNU Parallel or in a distributed environment using MPI, with both backends producing unified benchmarking outputs.}}

\label{fig:bench_frame}
\end{figure}

\begin{table}[htbp]
\caption{Algorithms used for Benchmarking.}
\label{tab:hpc-ml}
\centering
\setlength\tabcolsep{5pt}
\begin{tabular}{m{2.7cm} m{2.8cm} m{8cm}}
\toprule
\textbf{Benchmark} & \textbf{Domain} & \textbf{Characteristics} \\
\midrule
Back Propagation \citep{5306797}
    & Machine Learning
    & Neural network training via back-propagation. Compute-intensive, unstructured grid operations. \\
HotSpot \citep{5306797}
    & Physics Simulation
    & 2D thermal simulation for chip design. Compute-intensive, structured grid, stencil-based computations. \\
ParticleFilter \citep{5306797}
    & Medical Imaging
    & Probabilistic object tracking in videos. Irregular memory access, compute-heavy, suitable for parallelization. \\
SRAD \citep{5306797}
    & Image Processing
    & Noise reduction in images while preserving edges. Structured grid, compute-intensive, iterative. \\
LU Factorization \cite{golub2013matrix}
    & Linear Algebra
    & Matrix decomposition via Gaussian elimination for solving linear systems. Our benchmarks include dense LU and sparse LU.  
    \\
\bottomrule
\end{tabular}
\end{table}

In this paper, we utilize the results from our proposed benchmarking tool to analyze the mixed-precision profiles of classic algorithms in numerical simulations.  
Our simulation tasks as well as their categorization are presented in \tablename~\ref{tab:hpc-ml}. 
PROMISE is run  on algorithms from the Rodinia 
benchmark suite\footnote{https://rodinia.cs.virginia.edu/} \citep{5306797} 
and on well-established linear solving methods.  The following outlines the objectives and output measurement approaches for algorithms evaluated under the PROMISE simulation framework. 

\begin{itemize}
\item \textbf{Rodinia benchmark}: 
We test four programs: Back Propagation (neural network training), HotSpot (thermal simulation), Particle Filter (Monte Carlo localization), and SRAD (speckle-reducing anisotropic diffusion) from the Rodinia benchmark suite. 
Among these, 
we specify for which output array the desired accuracy
is achieved using \texttt{PROMISE\_CHECK\_ARRAY}: the 
error gradients of the output layer in the neural network for Back Propagation,  the temperature values of a grid after a transient thermal simulation for HotSpot, the x-coordinates of particles after filtering for ParticleFilter, and the final denoised image intensities after all diffusion iterations for SRAD.

\item \textbf{Linear system solver}:  
We solve the linear system $A x = b$ using LU factorization with pivoting
for both dense and sparse matrices. Dense systems involve compute-intensive operations with regular memory access, whereas sparse systems exhibit irregular access patterns, fill-in, and pivoting overhead. 

The solution process consists of:  
(i) factorizing the matrix into $L$ and $U$; (ii) forward substitution to solve $L y = b$; and (iii) backward substitution to solve $U x = y$. To improve numerical stability and avoid division by zero, pivoting is applied, resulting in a factorization of the form $P A = L U$, where $P$ is a permutation matrix. In the sparse case, Reverse Cuthill--McKee (RCM) ordering is additionally used to reduce fill-in and improve cache locality. 
For the dense solver, we use a synthetically generated matrix with a condition number of $10^4$. For the sparse solver, we use the \textit{sherman1} matrix from the SuiteSparse Matrix Collection~\cite{davis2011university}, which has a condition number of $1.56 \times 10^4$. In both cases, the ground-truth solution $x$ is set to the all-ones vector, and $b$ is computed as $b = A x$. The algorithm outputs the computed solution $x$, on which we perform the PROMISE check as an array-level validation.
  
\end{itemize}

\subsection{Empirical results and analysis}\label{sec:results}

\begin{figure}[htp]
\centering
\subfloat{\includegraphics[width=0.75\linewidth]{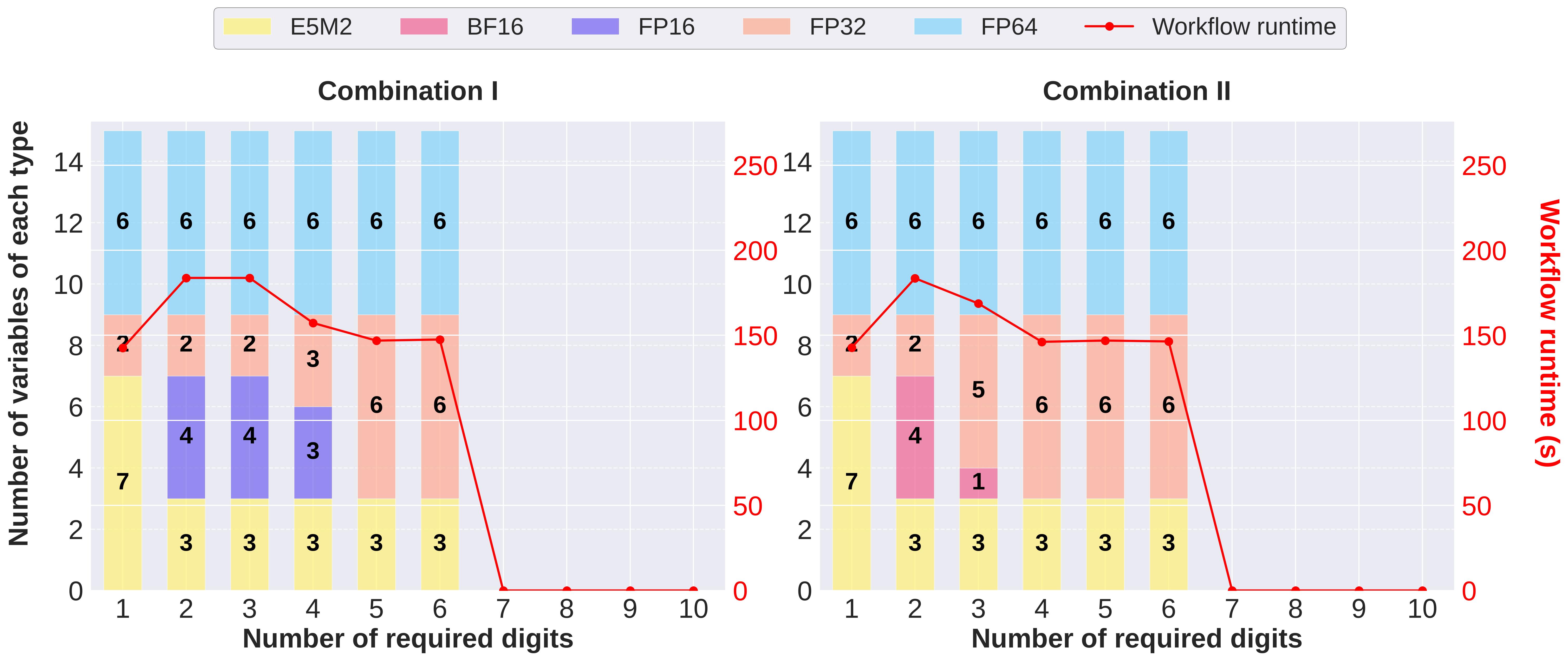}}

\caption{Precision configurations for Backprop with varying requested accuracies.}
\label{fig:bp}
\end{figure}

\begin{figure}[htp]
\centering
\subfloat{\includegraphics[width=0.75\linewidth]{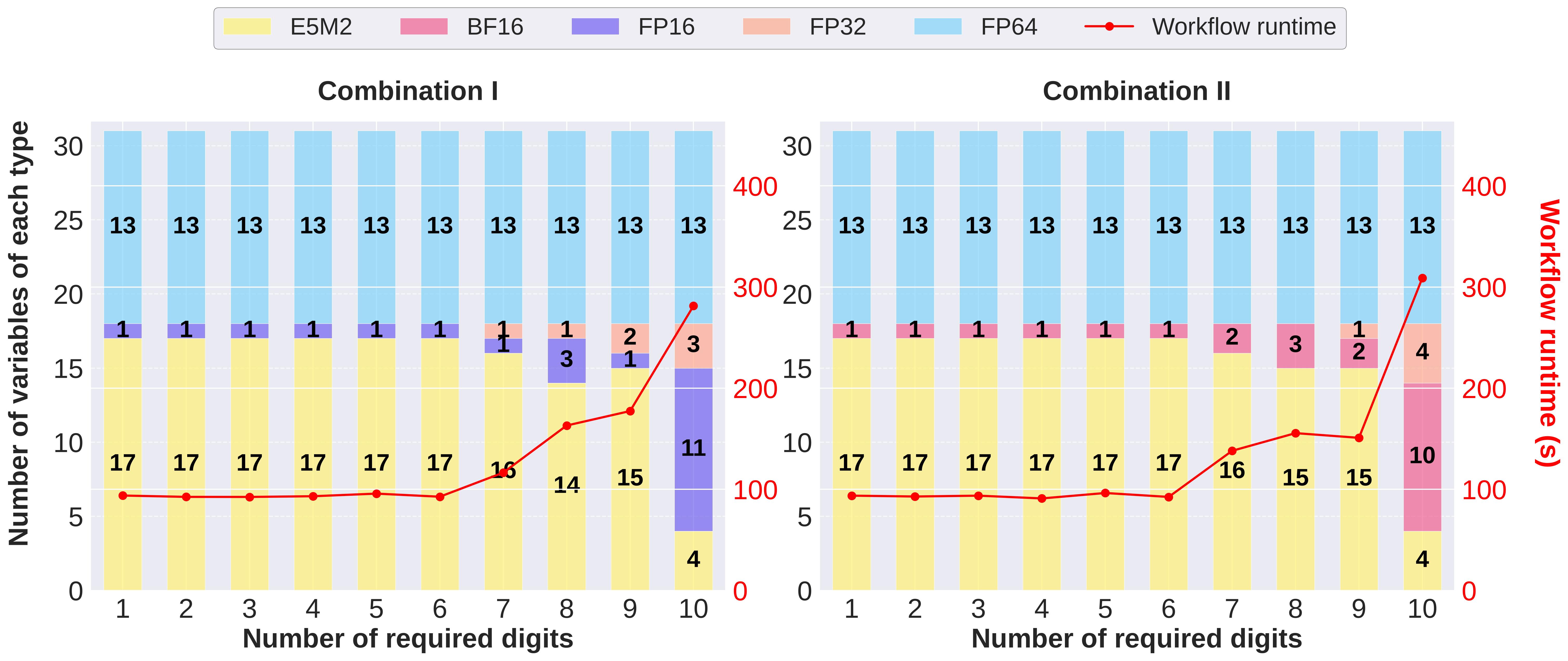}}\\
\caption{Precision configurations for Hotspot with varying requested accuracies.}
\label{fig:hotspot}
\end{figure}

\begin{figure}[htp]
\centering
\subfloat{\includegraphics[width=0.75\linewidth]{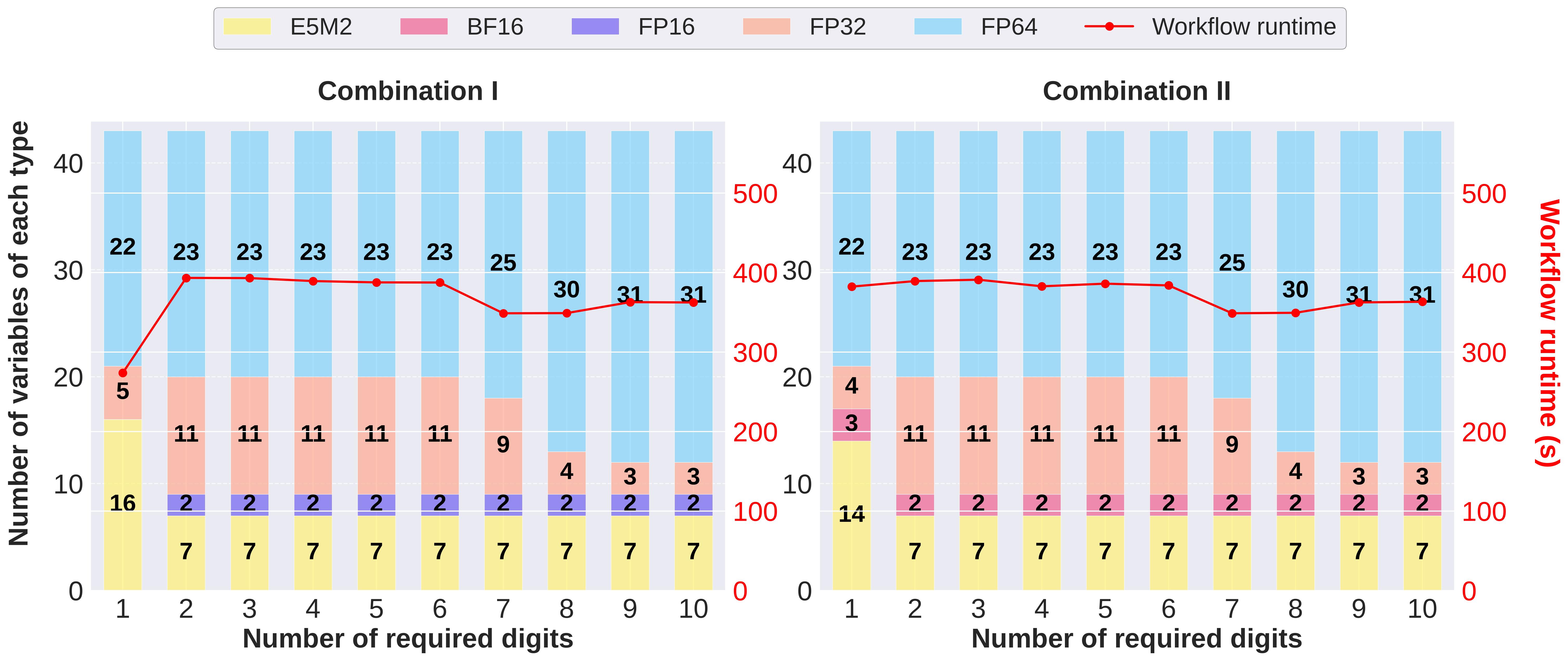}}
\caption{Precision configurations for Particle Filter with varying requested accuracies.}
\label{fig:pf}
\end{figure}

\begin{figure}[htp]
\centering
\subfloat{\includegraphics[width=0.75\linewidth]{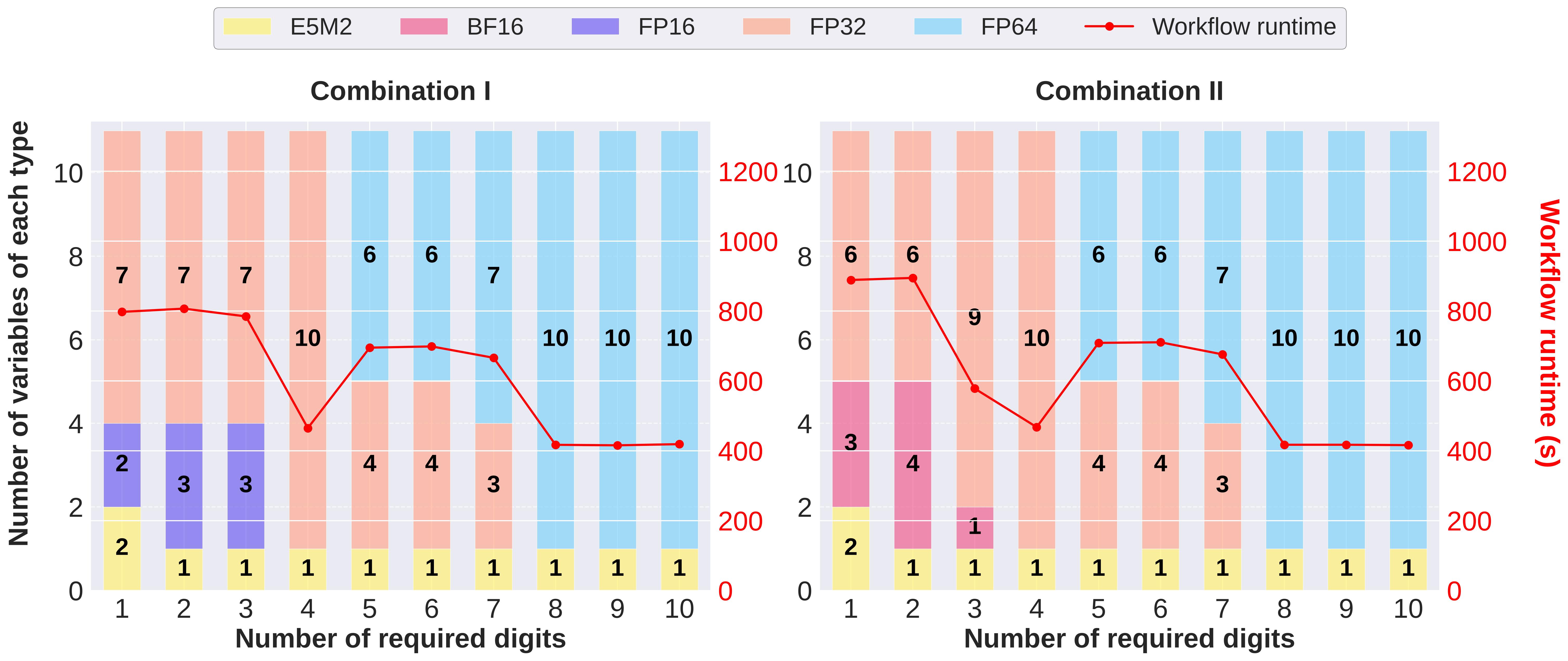}}\\
\caption{Precision configurations for SRAD with varying requested accuracies.}
\label{fig:srad}
\end{figure}

For the algorithms presented in \tablename~\ref{tab:hpc-ml},
PROMISE is evaluated across two precision combinations (I: E5M2-FP16-FP32-FP64; II: E5M2-BF16-FP32-FP64) to examine the number of variables of each type in the transformed codes. 
For a number of required digits in the results ranging from 1 to 10,
the type configurations obtained are shown in Figures~\ref{fig:bp} to \ref{fig:lu_all}. 

These figures also report the PROMISE workflow runtime for each requested accuracy, including reference computation and the repeated compilation and execution of candidate configurations during delta debugging. As illustrated in \figurename~\ref{fig:deltadebugging}, delta debugging does not exhaustively explore the search space. Consequently, the number of configurations evaluated, and hence the workflow runtime, can vary with the requested accuracy.
%
{The benchmark implementations consist of C/C++ numerical kernels with relatively small codebases. For completeness, }
the algorithms listed in \tablename~\ref{tab:hpc-ml}, together with results for additional precision combinations, are provided at 
\url{https://perso.lip6.fr/Fabienne.Jezequel/PROMISE_tests.pdf}. 
Readers are encouraged to refer to these algorithms throughout the following analysis. 

\subsubsection{Rodinia simulations}

\paragraph{Backprop}\label{sec:backprop_precision_migration}

{
In the Backprop kernel, which performs neural network training via backpropagation, error gradients achieve at most 6 correct digits, even in FP64, due to rounding and cancellation errors. The main numerical sensitivities arise from long dot-product reductions and the nonlinear sigmoid activation. Precision upgrades therefore target the sigmoid activation path first, followed by error-delta computations and, for higher accuracy requirements, weight-update accumulators.
}

In both combinations, the low-digit range (1 or 2 required digits) sees the first major transition: the sigmoid input path and its pre-activation accumulator are promoted from E5M2 to FP16 (in Combination~I) or to BF16 (in Combination~II) to stabilize the exponential evaluation near the steep region of the sigmoid. Once the critical regions of the sigmoid function are protected{---}by promoting the pre-activation sums and the activation path itself{---}the remaining error budget is dominated by the long summation reductions in the forward and backward passes. These reductions can still be performed in FP16 or BF16. 
For 5 and 6 required digits, selected critical reduction paths are promoted to FP32 to meet the higher accuracy target. 
Notably, 6 critical variables, mainly associated with the main weight matrices and their accumulators, must remain in full FP64 precision across all accuracy levels and precision combinations.


\paragraph{Hotspot}

Used for thermal simulation,
HotSpot is numerically dominated by an explicit stencil update that repeatedly computes a local temperature increment $\Delta$---a scaled sum of the power term, the discrete Laplacian in each direction, and the ambient coupling. This structure creates two distinct precision sensitivities: the per-cell update path exercised on every timestep ({which accumulates rounding errors over many iterations}) and the coefficient-preconditioning path that builds the geometry- and material-dependent scalars $C$ (capacitance), $R_x$, $R_y$, $R_z$ (thermal resistance), and their reciprocals $C_1$, $R_{x1}$, $R_{y1}$, $R_{z1}$. 

{
For both combinations, the type configurations are identical 
for 1 to 6
required digits: 
the scalars  $\Delta$, $C_1$, $R_{x1}$, $R_{y1}$, and $R_{z1}$
remain in E5M2 while the cell capacitance $C$ 
is promoted to FP16 or BF16. 
Since $C$ scales every cell update, its coarse quantization introduces a systematic bias. For 7 to 10 digits, additional thermal coefficients are promoted to FP32, indicating a shift from local rounding to accumulated scaling errors. PROMISE keeps the temperature arrays in FP64 while reducing the precision of the thermal coefficients and local update terms whenever possible. At higher accuracy requirements, both update-scaling factors and the neighbor differences defining the Laplacians require increased precision.
}


\paragraph{Particle Filter}
{
The Particle Filter kernel, 
 used for Monte Carlo localization in medical imaging, exhibits mixed-precision sensitivity driven mainly by control-flow decisions rather than long reduction chains. Fragility stems from nonlinear stochastic proposals, squared-residual observations, and exponential reweighting followed by weight normalization and Cumulative Distribution Function (CDF) construction. Since resampling compares uniform samples with the CDF, small errors can alter particle ancestry and redirect trajectories. Precision is therefore concentrated on nonlinear operations and the normalization/resampling boundary rather than applied uniformly.
}

For both combinations, the first major upgrade occurs when moving from 1 to 2 required correct digits: the random-number generation and nonlinear transforms (Gaussian proposal and exponential reweighting) shift from E5M2 to 16-bit or 32-bit precision 
to avoid bias in distribution tails and weight collapse. The configuration then remains unchanged from 2 to 6 required digits.  
Only beyond this range does the next transition become necessary; the weight summation and normalization step require higher precision (FP64) 
because relative errors propagate coherently through the CDF and affect all subsequent resampling decisions.

\paragraph{SRAD}

{
Used for noise reduction in images, the SRAD kernel exhibits mixed-precision sensitivity closely tied to its nonlinear diffusion pathway. The algorithm exponentially maps the input into a working image $J$ and computes global Region of Interest (ROI) statistics to estimate the baseline speckle noise level $q_0^2$. For each pixel, normalized directional measures based on the center pixel value $J_c$ are used to compute the local coefficient of variation $q^2$ and diffusion coefficient $c$. Because $c$ depends nonlinearly on $(q^2-q_0^2)$ and is clamped to $[0,1]$, rounding errors in the exponential mapping, ROI accumulation, or local normalizations can trigger saturation. Repeated image updates propagate these decisions, making the diffusion-coefficient computation the dominant numerical bottleneck.
}

In both type combinations, most of the precision challenges appear in the coefficient pathway. This includes 
$J_c$ normalization, forming the normalized directional measures, 
and the later rational and nonlinear steps that lead to $q^2$ and $c$. When only 1 or 2 correct digits are needed, the coefficient pathway can use the lowest-precision formats. As a result, divisions by $J_c$ and the squaring or accumulation steps are affected by rough quantization, which often pushes $c$ close to its saturation limits after clamping. 
When more digits are required, the formats move to higher precision (FP32 or FP64),
which greatly reduces these saturation effects.

\subsubsection{LU factorization}

\paragraph{Dense LU}

\begin{figure}[t]
\centering
\subfloat[Dense LU]{
    \includegraphics[width=0.75\linewidth]{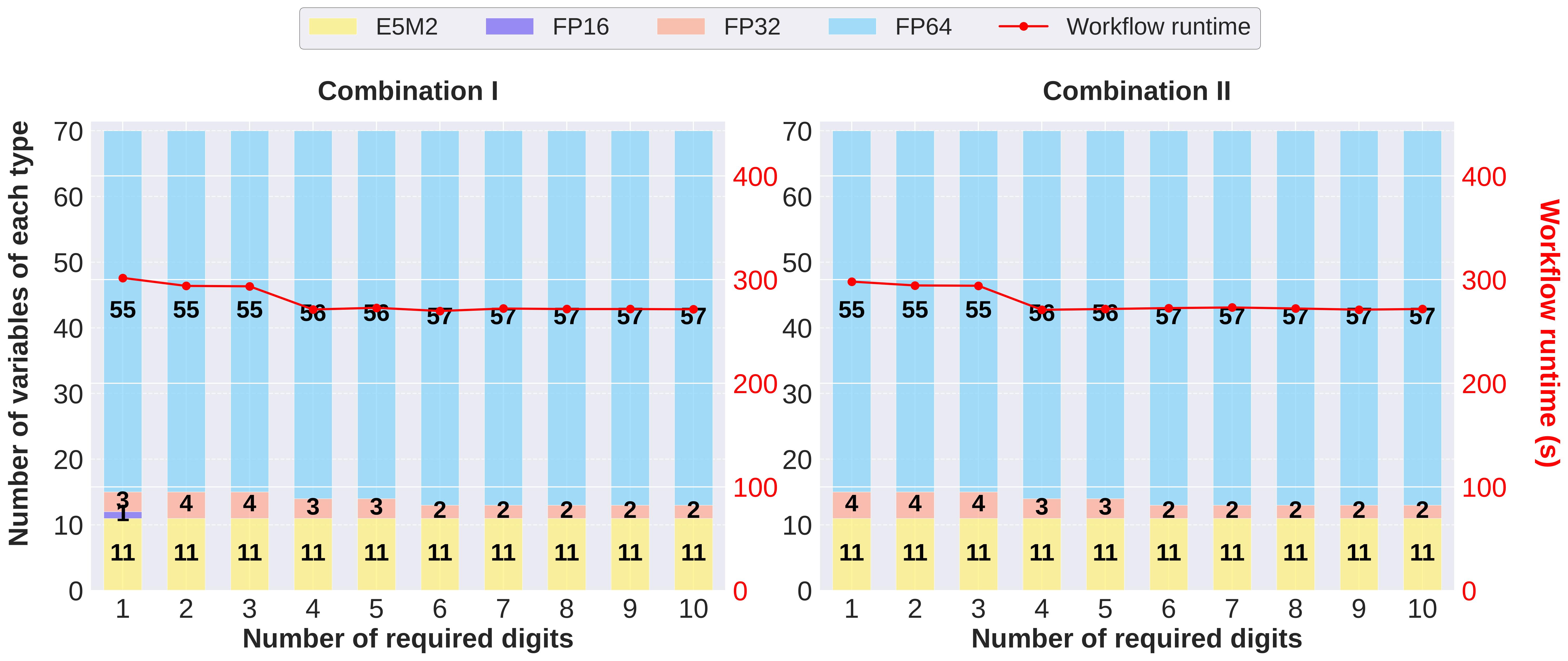}
}

\vspace{4pt} 
\subfloat[Sparse LU]{
    \includegraphics[width=0.75\linewidth]{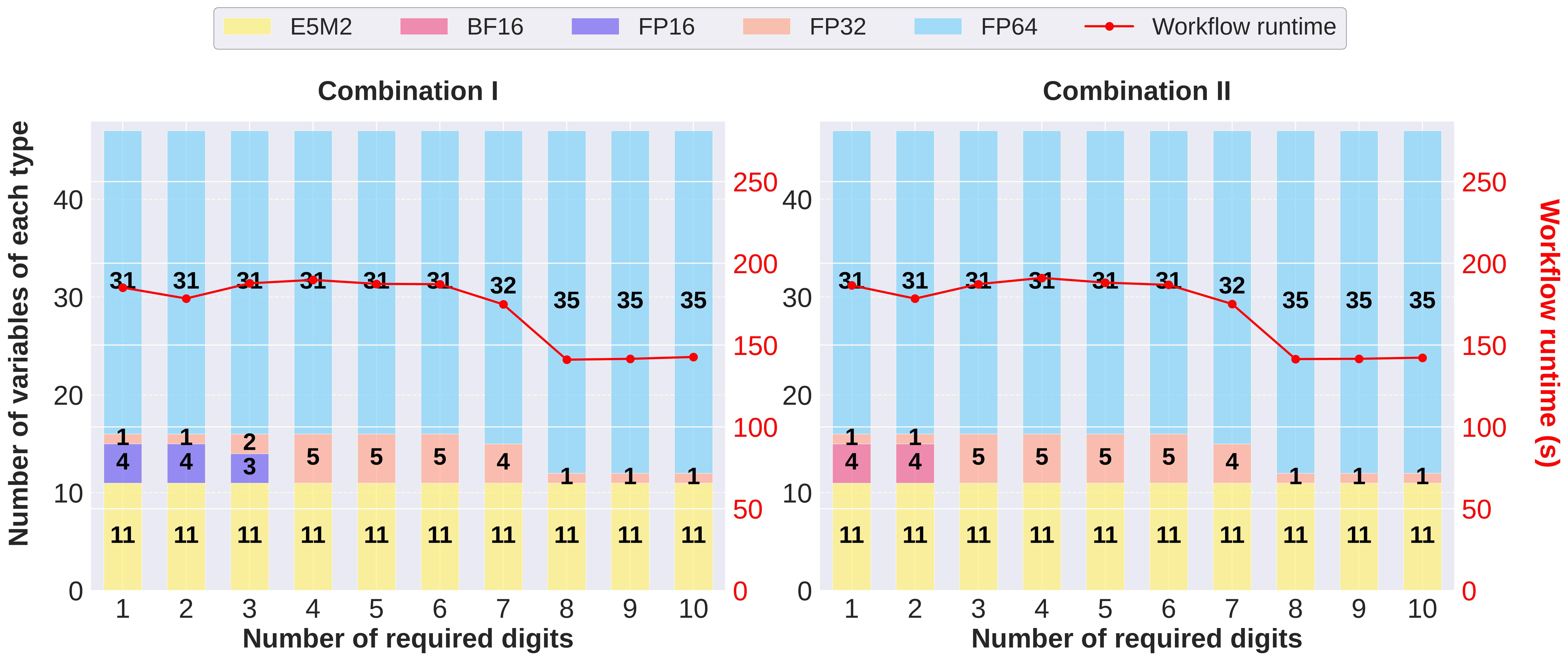}
}

\caption{Precision configurations for LU factorization (dense and sparse) under varying requested accuracies.}
\label{fig:lu_all}
\end{figure}

As shown in \figurename~\ref{fig:lu_all},
whatever the number of required digits, the type configurations obtained 
vary only slightly. 
For any required accuracy, the 8-bit format E5M2, which is mostly used for pivoting, consistently represents more than 20\% of the variables. 
Only in the lowest accuracy target of 
Combination I does the pivot divisor use FP16 (with the multiplier in FP32); all other cases use FP32 or FP64 for these scalars. FP32 appears for 2 to 4 variables, mainly the accumulator in the norm calculation and selected elements of the upper-triangular matrix. 
As a remark, the matrix used here
has a condition number of $10^4$. However, similar results have been obtained with matrices with a condition number of~$10$.

\paragraph{Sparse LU}
Similarly to the dense LU implementation, the sparse LU exhibits a relatively smooth precision transition as the number of required digits increases. 
The number of variables kept in E5M2, primarily those involved in  
pivoting decisions, remains constant across all required accuracy levels. 
From 8 required digits onward, most temporaries and scaling factors are promoted to FP64.  
The accumulator in the norm calculation consistently remains in FP32, while a few inner casts in the sort routine and forward substitution still use E5M2. 

In back substitution, the precision profile increases steadily with accuracy requirements: 
16-bit formats are sufficient for 1 to 3 required digits, FP32 (with FP64 used for the accumulator) covers the intermediate range up to 7 digits, and FP64 is adopted from 8 digits onward to guarantee stability across the multiple divisions performed per row.

\paragraph{Discussion}

Averaged over all benchmarks and target accuracy levels, 43\% of the variables can be represented with 32-bit precision or less, and
    28\% of the variables can be reduced to 16-bit precision or lower.

These results  indicate that FP64 is systematically over-provisioned for the levels of accuracy considered. 
The use of FP32 tends to align with a small number of sensitivity bottlenecks (e.g., exponentials/logs, variance-like reductions, and weight/CDF accumulation paths).
Customized precision enables one to find suitable type configurations
that include low precision formats available on the computing target. 
In particular, on average E5M2 represents 23\% of the
variables in the configurations obtained.
As a remark, using E4M3 instead leads to a
similar proportion, as shown in additional results
presented in \url{https://perso.lip6.fr/Fabienne.Jezequel/PROMISE_tests.pdf}. 

\subsection{One-bit exploration}

\begin{figure}[t]
\centering

\subfloat[Hotspot]{
    \includegraphics[width=0.7\linewidth]{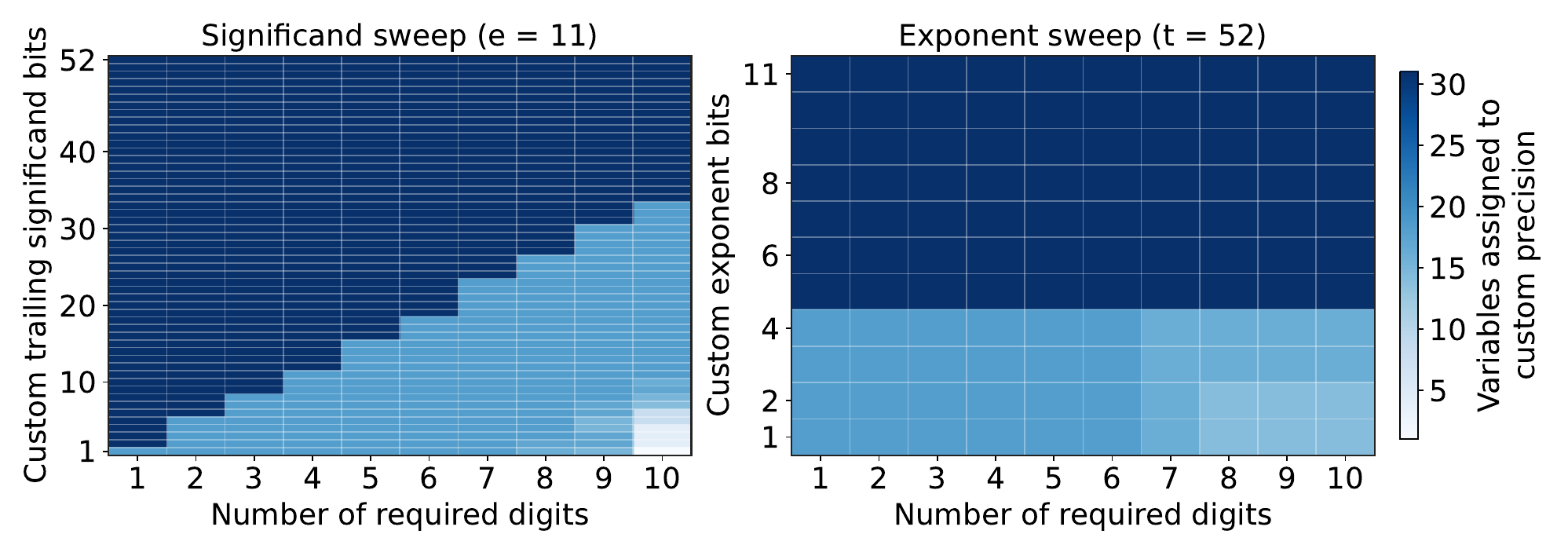}
}

\subfloat[Dense LU]{
    \includegraphics[width=0.7\linewidth]{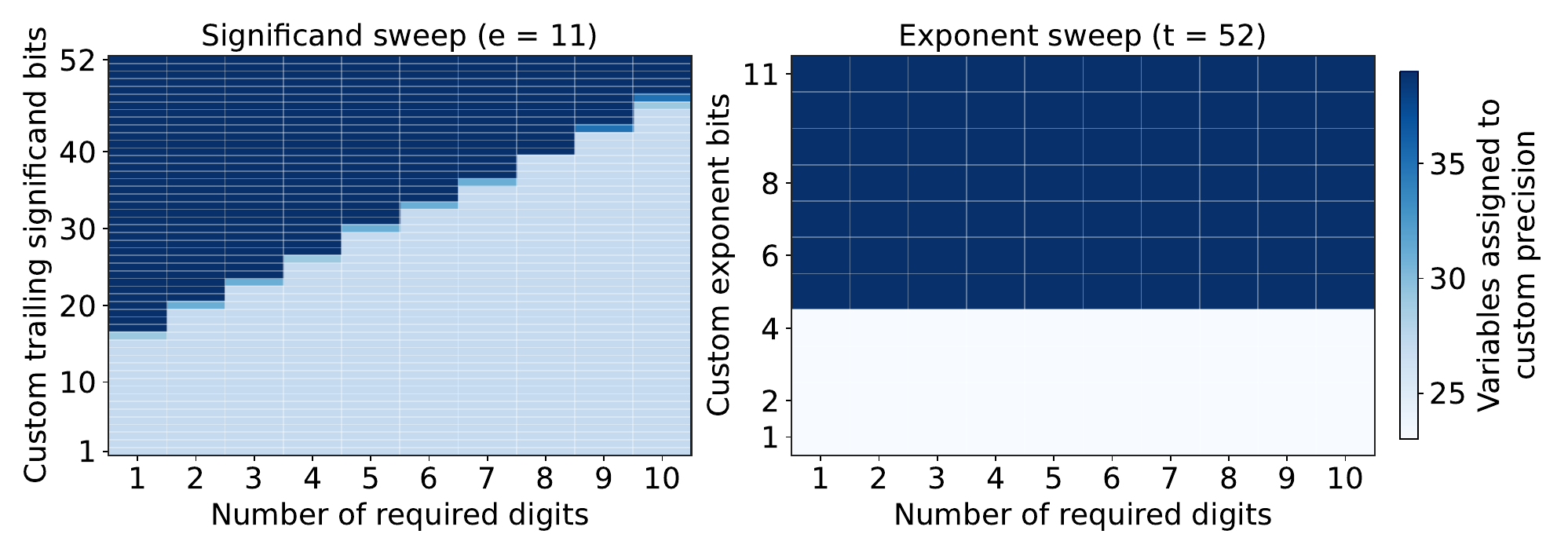}
}

\caption{{One-bit granularity exploration for Hotspot and dense LU. The $x$-axis reports the number of required digits, and the $y$-axis reports the swept custom bitwidth. In each subfigure, the left panel sweeps the trailing-significand bitwidth with the exponent bitwidth fixed to 11, whereas the right panel sweeps the exponent bitwidth with the trailing-significand bitwidth fixed to 52. Color indicates the number of variables assigned by PROMISE to the tested custom precision.}}
\label{fig:1-bit-exp}
\end{figure}

To directly assess the customized-precision capability of PROMISE, an 
experiment at one-bit granularity is conducted on two representative kernels from \tablename~\ref{tab:hpc-ml}: Hotspot and dense LU factorization. They capture two contrasting numerical structures, namely a regular stencil-based Rodinia kernel and a direct linear-algebra solver. 
Such fine-grained precision exploration may be useful for highly customizable FPGA applications. The purpose of this experiment is not to propose these experimental configurations as practical hardware types, but to isolate the respective effects of exponent range and significand precision on the number of correct digits delivered by the application.
For each required accuracy, PROMISE is run with two types: FP64 and a custom type. 
In the custom type, 
either the exponent size is fixed to 11 bits while the trailing significand size varies from 1 to 52 bits, or the trailing significand size is fixed to 52 bits while the exponent size varies from 1 to 11 bits.
\figurename~\ref{fig:1-bit-exp} presents the number of variables assigned to a custom type, out of a total of 31 variables for Hotspot and 70 variables for dense LU.

{
For both dense LU and Hotspot, the exponent sweep saturates quickly: once the exponent field reaches about five bits, additional exponent bits do not substantially increase the number of variables assigned to the custom format. The significand sweep instead shifts steadily toward larger trailing significand sizes as the requested number of significant digits increases. This shift is stronger for dense LU, whose pivoting and elimination updates amplify accumulated rounding errors, whereas Hotspot shows a smoother progression consistent with a regular stencil computation.}

\subsection{Performance validation on GPU}

{
We further evaluate three representative benchmarks on GPU: 
Backprop, Hotspot, and dense LU. These kernels exhibit different computational characteristics: 
Backprop contains neural-network forward/backward propagation, Hotspot is a regular stencil computation with uniform memory access, and dense LU is dominated by pivoting and dependent elimination updates.}
{
Since the current version of PROMISE runs on CPU, the type configurations mentioned in~\ref{sec:results} are obtained on CPU for Combinations I and II. For each benchmark and required accuracy, we then evaluate on GPU the performance obtained for both combinations. A GPU extension of PROMISE remains a future direction. 

In~\ref{sec:results}, the following inputs are used:
65,536 
elements for Backprop, a $512 \times 512$ Hotspot grid for two iterations, and a $500 \times 500$ dense matrix 
for LU. 
Here, the 
problem sizes are 262,144 input elements for Backprop, a $5000 \times 5000$ dense matrix for LU, and the same $512 \times 512$ Hotspot grid for two iterations. 
Thus, we assess the GPU performance of the PROMISE-derived configurations, with enlarged inputs for some benchmarks. 
}

\begin{figure}
\centering
\subfloat{\includegraphics[width=0.85\linewidth]{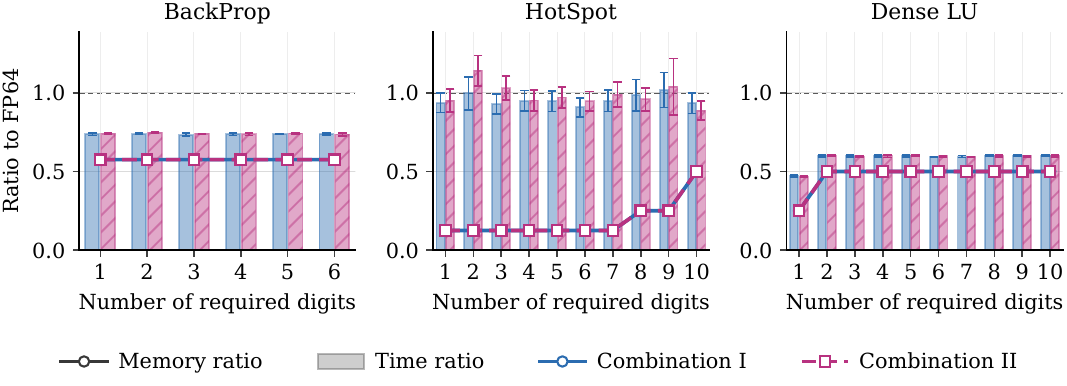}}

\caption{Evaluation of PROMISE-derived mixed-precision CUDA programs for the two precision combinations. Ratios are normalized by the FP64 CUDA baseline, and lower values indicate lower tracked device allocation or faster execution. Memory allocation is measured from tracked \texttt{cudaMalloc} byte counts. Error bars represent one propagated standard deviation of the time ratio, accounting for variability in both the mixed-precision and FP64 timings over three measured runs each.}
\label{fig:gpu_acceleration}
\end{figure}

{
The runs were performed on one NVIDIA H100 GPU on the Jean Zay supercomputer at IDRIS, with CUDA code compiled for the \texttt{sm\_90} architecture.
The CUDA ports preserve the PROMISE-selected reduced-precision assignments and map the formats from Combinations I and II to H100-supported 
types: FP64 to \texttt{double}, FP32 to \texttt{float}, FP16 (\texttt{flx::floatx<5,10>}) to \texttt{\_\_half}, BF16 (\texttt{flx::floatx<8,7>}) to \texttt{\_\_nv\_bfloat16}, and E5M2 (\texttt{flx::floatx<5,2>}) to CUDA FP8 storage. The implementations are direct CUDA ports of the PROMISE-derived programs, not cuBLAS/cuDNN or WMMA/MMA tensor-core rewrites. 
Thus, the timings presented in \figurename~\ref{fig:gpu_acceleration} 
measure the effect of applying PROMISE-derived precision assignments to CUDA kernels, rather than the peak tensor-core throughput of H100. For each type configuration, execution time is averaged over three measured runs after one untimed warm-up. 
Figure~\ref{fig:gpu_acceleration} reports the execution time and memory consumption of the mixed-precision CUDA versions relative to FP64. Memory consumption is measured from the bytes allocated through \texttt{cudaMalloc} for device-resident arrays and work buffers. It therefore represents the allocation footprint rather than the number of source-level variables.

}

{\figurename~\ref{fig:gpu_acceleration} shows that the  impact of mixed precision strongly depends on the kernel structure. Backprop reduces the tracked device allocation to about $57.5\%$ of the FP64 baseline and obtains stable time ratios between $0.734$ and $0.745$, corresponding to a speedup of about $1.34$--$1.36\times$. 
Its memory ratio remains almost unchanged across digit targets and the two type combinations, despite differences in the types used in the source code.
The memory and time ratios remain similar because the dominant arrays retain the same storage formats, while the small precision changes have little impact on GPU execution.
}

{
Hotspot shows the largest memory 
reduction. 
Its memory ratio is $0.125$ for 1--7 required digits, increases to $0.25$ for 8--9 digits, and reaches $0.5$ for 10 digits because the main temperature, power, and result grids are stored in the PROMISE-selected field type. 
The apparent discrepancy with Fig.~\ref{fig:hotspot} 
is explained by the types used for scalar coefficients and register-level update quantities, which do not affect the storage width of the three dominant grids.
Moreover, FP16 and BF16 both occupy two bytes, so differences between Combination~I and Combination~II do not necessarily change the memory ratio. 
The time ratio ranges from about $0.89$ to $1.14$. 
Timing overhead can be observed because the direct CUDA implementation stores data in low precision but converts values to higher precision for scalar arithmetic within the stencil update.
}

{
For dense LU, a one-digit accuracy requirement reduces memory allocation to 25\% of FP64, compared with about 50\% for the other tested accuracy levels.
This transition reflects the 
type of the dominant packed matrix in the CUDA implementation: FP16-sized for one digit and FP32-sized from two digits onward. 
Memory ratio is governed by the $n^2$ matrix storage, not by the number of 
variables previously reported; 
many variables counted as FP64 in Fig.~\ref{fig:lu_all} are scalars, temporaries, or control quantities and do not dominate device allocation. 
Dense LU obtains the largest timing improvement. 
The time ratio ranges from about $0.47$ to $0.60$, or about $1.66$--$2.14\times$ speedup. 
}

Overall, the GPU results confirm that PROMISE-derived precision assignments substantially reduce device memory requirements, potentially allowing simulations to fit on fewer GPUs or cluster nodes and reducing MPI overhead when GPU memory capacity is limiting. Runtime gains depend on whether the selected variables dominate memory allocation and whether low-precision storage maps efficiently to the benchmark’s SIMT execution pattern. 


\section{Conclusion and perspectives}\label{sec:conclude}
In this paper, we advance PROMISE by incorporating customized-precision emulation and enabling scalable deployment. This allows users to exploit more reduced-precision variables 
within a limited time frame, thereby promoting broader adoption of mixed-precision. 
Leveraging PROMISE, we perform simulations on our benchmarks 
to identify mixed-precision configurations for direct linear solvers and Rodinia floating-point kernels. 
Our empirical results show that PROMISE enables substantial precision reductions while meeting the required accuracy, which can in turn lead to performance improvements. 
The extension of PROMISE to GPUs is a planned perspective that 
would benefit from a GPU version of CADNA. 
Future work will also explore support for distinct storage and
computation precisions, such as bfloat16 storage with fp32
computation, to exploit GPU tensor cores and AI accelerators. We will
further investigate precision autotuning for iterative solvers, with
particular attention to potential increases in iteration count when
low precision is applied to automatically selected variables.

\bibliographystyle{ACM-Reference-Format}
\bibliography{refs}

@article{GRAILLAT2019101017,
title = {Auto-tuning for floating-point precision with Discrete Stochastic Arithmetic},
journal = {J. Comput. Sci.},
volume = {36},
year = {2019},
doi = {10.1016/j.jocs.2019.07.004},
author = {Stef Graillat and Fabienne Jézéquel and Romain Picot and François Févotte and Bruno Lathuilière},
}

@article{Fasi2023CPFloat,
author = {Fasi, Massimiliano and Mikaitis, Mantas},
title = {{CPFloat}: A C Library for Simulating Low-precision Arithmetic},
journal = {ACM Trans. Math. Softw.},
volume = {49},
number = {2},
articleno = {18},
pages = {1--32},
year = {2023},
doi = {10.1145/3585515},
publisher = {Association for Computing Machinery}
}

@article{Vignes2004,
  author = {Vignes, Jean},
  title = {Discrete stochastic arithmetic for validating results of numerical software},
  journal = {Numer. Algorithms},
  year = {2004},
  volume = {37},
  number = {1--4},
  pages = {377--390},
  doi = {10.1023/B:NUMA.0000049483.75679.ce},
}

@INPROCEEDINGS{9251260,
  author={Parasyris, Konstantinos and Laguna, Ignacio and Menon, Harshitha and Schordan, Markus and Osei-Kuffuor, Daniel and Georgakoudis, Giorgis and Lam, Michael O. and Vanderbruggen, Tristan},
  booktitle={IEEE International Symposium on Workload Characterization}, 
  title={{HPC-MixPBench}: An HPC Benchmark Suite for Mixed-Precision Analysis}, 
  year={2020},
  volume={},
  number={},
  pages={25--36},
  doi={10.1109/IISWC50251.2020.00012}
}

@article{Jez2008,
  author = {J{\'e}z{\'e}quel, F. and Chesneaux, J.-M.},
  title = {{CADNA}: A Library for Estimating Round-Off Error Propagation},
  journal = {Computer Physics Communications},
  volume = {178},
  number = {12},
  pages = {933--955},
  year = {2008},
  doi = {10.1016/j.cpc.2008.02.013}
}

@article{kashi2025,
  author = {Kashi, Aditya and Lu, Hao and Brewer, Wesley and Rogers, David and Matheson, Michael and Shankar, Mallikarjun and Wang, Feiyi},
  title = {Mixed-precision numerics in scientific applications: survey and perspectives},
  journal = {J. Supercomput.},
  year = {2026},
  volume = {82},
  number = {5},
  doi = {10.1007/s11227-026-08264-4},
}

@article{Higham_Mary_2022, title={Mixed precision algorithms in numerical linear algebra}, volume={31}, DOI={10.1017/S0962492922000022}, journal={Acta Numer.}, 
author={Higham, Nicholas J. and Mary, Theo}, 
year={2022}, 
pages={347–414}}

@article{10.1145/3368086,
author = {Flegar, Goran and Scheidegger, Florian and Novakovi\'{c}, Vedran and Mariani, Giovani and Tom\'{a}s, Andr\'{e}s E. and Malossi, A. Cristiano I. and Quintana-Ort\'{\i}, Enrique S.},
title = {{FloatX}: A {C++} Library for Customized Floating-Point Arithmetic},
year = {2019},
publisher = {ACM},
address = {USA},
volume = {45},
number = {4},
doi = {10.1145/3368086},
journal = {ACM Trans. Math. Softw.},
articleno = {40},
numpages = {23},
}

@INPROCEEDINGS{Rubio2016,
  author={Rubio-González, Cindy and Nguyen, Cuong and Mehne, Benjamin and Sen, Koushik and Demmel, James and Kahan, William and Iancu, Costin and Lavrijsen, Wim and Bailey, David H. and Hough, David},
  booktitle={IEEE/ACM 38th International Conference on Software Engineering}, 
  title={Floating-Point Precision Tuning Using Blame Analysis}, 
  year={2016},
  volume={},
  number={},
  pages={1074--1085},
  doi={10.1145/2884781.2884850}}

@article{10.1177/1094342016652462,
author = {Lam, Michael O and Hollingsworth, Jeffrey K},
title = {Fine-grained floating-point precision analysis},
year = {2018},
publisher = {Sage Publications, Inc.},
address = {USA},
volume = {32},
number = {2},
doi = {10.1177/1094342016652462},
journal = {Int. J. High Perform. Comput. Appl.},
pages = {231--245},
numpages = {15}
}

@InProceedings{10.1007/978-3-030-72654-6_29,
author="J{\'e}z{\'e}quel, Fabienne
and Hoseininasab, Sara sadat
and Hilaire, Thibault",
title="Numerical Validation of Half Precision Simulations",
booktitle="Trends and Applications in Information Systems and Technologies",
year="2021",
publisher="Springer International Publishing",
address="Cham",
pages="298--307",
}

@article{10.1145/1236463.1236468,
  author = {Fousse, Laurent and Hanrot, Guillaume and Lef{\`e}vre, Vincent and P{\'e}lissier, Patrick and Zimmermann, Paul},
  title = {{MPFR}: a multiple-precision binary floating-point library with correct rounding},
  journal = {ACM Trans. Math. Softw.},
  year = {2007},
  volume = {33},
  number = {2},
  numpages = {15},
  doi = {10.1145/1236463.1236468},
}

@INPROCEEDINGS{5306797,
  author={Che, Shuai and Boyer, Michael and Meng, Jiayuan and Tarjan, David and Sheaffer, Jeremy W. and Lee, Sang-Ha and Skadron, Kevin},
  booktitle={IEEE International Symposium on Workload Characterization}, 
  title={Rodinia: A benchmark suite for heterogeneous computing}, 
  year={2009},
  volume={},
  number={},
  pages={44--54},
  doi={10.1109/IISWC.2009.5306797}
}

@article{10.1145/3381039,
author = {Cherubin, Stefano and Agosta, Giovanni},
title = {Tools for Reduced Precision Computation: A Survey},
year = {2020},
publisher = {ACM},
address = {USA},
volume = {53},
number = {2},
doi = {10.1145/3381039},
journal = {ACM Comput. Surv.},
numpages = {35},
}

@INPROCEEDINGS{10820739,
  author={Vanover, Jackson and Altuntas, Alper and Rubio-González, Cindy},
  booktitle={SC24-W: Workshops of the International Conference for High Performance Computing, Networking, Storage and Analysis}, 
  title={Toward Automated Precision Tuning of Weather and Climate Models: A Case Study}, 
  year={2024},
  volume={},
  number={},
  pages={148--159},
  doi={10.1109/SCW63240.2024.00026}
}

@inproceedings{10.1145/2503210.2503296,
author = {Rubio-Gonz\'{a}lez, Cindy and Nguyen, Cuong and Nguyen, Hong Diep and Demmel, James and Kahan, William and Sen, Koushik and Bailey, David H. and Iancu, Costin and Hough, David},
title = {Precimonious: tuning assistant for floating-point precision},
year = {2013},
publisher = {ACM},
address = {USA},
doi = {10.1145/2503210.2503296},
booktitle = {Proceedings of the International Conference on High Performance Computing, Networking, Storage and Analysis},
numpages = {12},
series = {SC '13}
}

@article{VIGNES1993233,
title = {A stochastic arithmetic for reliable scientific computation},
journal = {Mathematics and Computers in Simulation},
volume = {35},
number = {3},
pages = {233--261},
year = {1993},
doi = {10.1016/0378-4754(93)90003-D},
author = {Vignes, Jean},
}

@inproceedings{10548640,
  author = {Wang, Yutong and Rubio-Gonz{\'a}lez, Cindy},
  title = {Predicting performance and accuracy of mixed-precision programs for precision tuning},
  booktitle = {IEEE/ACM 46th International Conference on Software Engineering},
  year = {2024},
  articleno = {15},
  pages = {1--13},
  numpages = {13},
  doi = {10.1145/3597503.3623338},
}

@ARTICLE{988498,
  author={Zeller, A. and Hildebrandt, R.},
  journal={IEEE Transactions on Software Engineering}, 
  title={Simplifying and isolating failure-inducing input}, 
  year={2002},
  volume={28},
  number={2},
  pages={183--200},
  doi={10.1109/32.988498}
}

@inproceedings{10.1145/3213846.3213862,
author = {Guo, Hui and Rubio-Gonz\'{a}lez, Cindy},
title = {Exploiting community structure for floating-point precision tuning},
year = {2018},
publisher = {ACM},
address = {USA},
doi = {10.1145/3213846.3213862},
booktitle = {Proceedings of the 27th ACM SIGSOFT International Symposium on Software Testing and Analysis},
pages = {333--343},
numpages = {11},
series = {ISSTA 2018}
}

@article{ieee754,
  author = {{IEEE Computer Society}},
  title = {IEEE Standard for Floating-Point Arithmetic},
  journal = {IEEE Std 754-2019},
  year = {2019},
  publisher = {IEEE},
  doi = {10.1109/IEEESTD.2019.8766229},
}

@article{Eberhart2015HighPerformance,
  author    = {P. Eberhart and J. Brajard and P. Fortin and F. J{\'e}z{\'e}quel},
  title     = {High Performance Numerical Validation using Stochastic Arithmetic},
  journal   = {Reliable Computing},
  volume    = {21},
  pages     = {35--52},
  year      = {2015}
}

@book{golub2013matrix,
  title     = {Matrix Computations},
  author    = {Golub, Gene H. and Van Loan, Charles F.},
  edition   = {4},
  year      = {2013},
  publisher = {Johns Hopkins University Press},
  address   = {Baltimore},
  isbn      = {978-1421407944}
}

@article{davis2011university,
  author = {Davis, Timothy A. and Hu, Yifan},
  title = {The University of Florida sparse matrix collection},
  journal = {ACM Trans. Math. Softw.},
  volume = {38},
  number = {1},
  pages = {1--25},
  year = {2011},
  doi = {10.1145/2049662.2049663}
}

@inproceedings{Menon2018ADAPT,
  author    = {Harshitha Menon and Michael O. Lam and Daniel Osei{-}Kuffuor and
               Markus Schordan and Scott Lloyd and Kathryn Mohror and
               Jeffrey Hittinger},
  title     = {{ADAPT}: Algorithmic Differentiation Applied to Floating-Point Precision Tuning},
  booktitle = {Proceedings of the International Conference for High Performance Computing,
               Networking, Storage, and Analysis},
  pages     = {48:1--48:13},
  year      = {2018},
  doi       = {10.1109/SC.2018.00051}
}

@inproceedings{Singh2023CHEFFP,
  author    = {Garima Singh and Baidyanath Kundu and Harshitha Menon and
               Alexander Penev and David J. Lange and Vassil Vassilev},
  title     = {Fast and Automatic Floating-Point Error Analysis with {CHEF-FP}},
  booktitle = {Proceedings of the IEEE International Parallel and Distributed Processing Symposium},
  pages     = {1018--1028},
  year      = {2023},
  doi       = {10.1109/IPDPS54959.2023.00105}
}

@article{carson2025,
author = {Carson, Erin and Chen, Xinye},
title = {pychop: Emulating Low-Precision Arithmetic in Numerical Methods and Neural Networks},
year = {2026},
publisher = {Association for Computing Machinery},
address = {New York, NY, USA},
issn = {0098-3500},
doi = {10.1145/3840284},
journal = {ACM Trans. Math. Softw.},
}

\end{document}